\documentclass[10pt]{article}

\usepackage{amsmath}
\usepackage{amssymb}
\usepackage[bbgreekl]{mathbbol}

\usepackage{graphicx}

\usepackage{cite}
\usepackage{soul}
\usepackage{color}

\usepackage[labelfont=bf,labelsep=period,justification=raggedright]{caption}

\makeatletter
\renewcommand{\@biblabel}[1]{\quad#1.}
\makeatother

\date{}

\usepackage{lscape}
\usepackage{caption}

\def \bfr{{\bf r}}
\def \XX{\mathbb{X}}
\def \YY{\mathbb{Y}}
\def \Dnx{\mathcal{D}_{nx}}

\def \Dist{\mathcal{D}}
\def \LN{L_N}
\def \LC{L_C}
\def \Qab{Q^{\alpha\beta}}

\begin{document}

\begin{flushleft}
  {\Large
    \textbf{Polymer uncrossing and knotting in protein folding, and their role
      in minimal folding pathways}
  }
  \\
  Ali R. Mohazab$^{1}$, 
  Steven S. Plotkin$^{1,\ast}$
  \\
  \bf{1} Department of Physics and Astronomy, University of British
  Columbia, Vancouver, B.C. Canada 
  \\
  $\ast$ E-mail: steve@physics.ubc.ca
\end{flushleft}

\section*{Abstract}
We introduce a method for calculating the extent to which chain
non-crossing is important in the most efficient, optimal trajectories
or pathways for a protein to fold. This involves recording all
unphysical crossing events of a ghost chain, and calculating the
minimal uncrossing cost that would have been required to avoid such
events. A depth-first tree search algorithm is applied to find minimal
transformations to fold $\alpha$, $\beta$, $\alpha/\beta$,  and
knotted proteins.  In all cases, the extra uncrossing/non-crossing
distance is a small fraction of the total distance travelled by a
ghost chain. Different structural classes may be distinguished by the
amount of extra uncrossing distance, and the effectiveness of such
discrimination is compared with other order parameters. It was seen
that non-crossing distance over chain length provided the best
discrimination between structural and kinetic classes. 
The scaling of non-crossing distance with chain length implies an
inevitable crossover to entanglement-dominated folding mechanisms for
sufficiently long chains. 
We further quantify the minimal folding pathways by collecting the
sequence of uncrossing moves, which generally involve leg, loop, and
elbow-like uncrossing moves, and rendering the collection of these
moves over the unfolded ensemble as a multiple-transformation ``alignment''.
The consensus minimal pathway is constructed and shown schematically
for representative cases of an $\alpha$, $\beta$, and knotted
protein. An overlap parameter is defined between pathways; we find
that $\alpha$ proteins have minimal overlap indicating diverse folding
pathways, knotted proteins are highly constrained to follow a dominant
pathway, and $\beta$ proteins are somewhere in between. Thus we have
shown how topological chain constraints can induce dominant pathway
mechanisms in protein folding.  

\section*{Author Summary}
Researchers have long focused on the problem of how to design and
predict low-energy protein structures from amino acid sequence,
without worrying very much about how those structures can be found
without the protein getting tangled up in its own game of Twister gone
awry. This problem becomes a
serious one for proteins whose folded structures form knots, of which
several hundred have now been found.  %%
Here, we develop and apply a formalism to find the way a protein would
fold up if it could do so with the least amount of motion. We know
proteins generally don't fold up the same way every
time. Nevertheless, one can't help but wonder if the constraints due 
to the presence of the protein chain itself could, in some cases, be
so severe that no matter where the protein started from, nearly a
single folding pathway would be induced, akin to navigating a maze. 
We found that the answer depends on the structure- for a typical
$\alpha$-helical protein there are many roads to Rome, while for a knotted
protein the solution is much more maze-like: chain non-crossing
constraints can induce a mechanism to folding, and a pathway to the
folded structure.

\section*{Introduction}
Protein folding is a structural transformation, from a
disordered-polymer conformational ensemble to an ordered, well-defined
structure. Quantifying
the dynamical mechanism by which this occurs has been a long-standing
problem of interest to both theorists and
experimentalists~\cite{WolynesPG92:spinb,Chan93,Wolynes95,Garel96,Dobson98,FershtBook00,EatonWA00,EnglanderSW00,Pande00RMP,Shea2001,PlotkinSS02:quartrev1,PlotkinSS02:quartrev2,SnowCD02,OlivebergM05rev,KhatibF11,Lindorff-Larsen11}.  
It is currently not possible experimentally to capture the full
dynamical mechanism of a folding protein in atomic detail, start to
finish. Photon counting analyses of single molecule folding
trajectories can now extract the mean transition path
time across the distribution of productive folding
pathways~\cite{ChungHS12}. Typically however,
snapshots of the participation of various residues in the folding
transition state are used to infer the relative importance of amino
acids in defining the protein folding
nucleus~\cite{Fersht92,AbkevichVI94,FershtAR95,DaggettV96,GianniS03,Klimov98,OlivebergM98,MartinezJC98,FershtBook00,ClementiC03jmb,EjtehadiMR04,OztopB04,BodenreiderC05,SosnickTR06,WensleyBG09}. An
idea of how the nucleus grows as folding proceeds may be gained by
exploring the native shift in the transition state as denaturant
concentration is increased~\cite{TernstromT99}, but ideally the goal is
to quantify folding mechanisms under constant environmental
conditions. To this end, simulations and theory have proved an
invaluable
tool~\cite{ChanHS90,Wolynes97cap,NymeyerH98:pnas,DuR98:jcp,Nymeyer99:PNAS,Shoemaker99b,ZhouY99,PlotkinSS00:pnas,PlotkinSS02:Tjcp,PlotkinSS02:quartrev2,SnowCD02,FavrinG03,WeiklTR07},
and have in many respects succeeded in reproducing the general
features of the folding pathway (see
e.g. references~\cite{MaityH05,WeinkamP05} 
for cytochrome c).

One %%
conceptual refinement to arise from theoretical and
simulation studies is the study of ``good'' reaction coordinates that
correlate with commitment probability to complete the protein folding
reaction~\cite{DellagoC98,DuR98:jcp,BolhuisPG02,HummerG04,BestRB05,MaraglianoL06,vanderVaartA07}.
Reaction coordinates must generally take into account the energy
surface on which the molecule of interest is undergoing conformational
diffusion~\cite{FischerS92,YangH07,BranduardiD07}, and the Markovian or
non-Markovian nature of the diffusion~\cite{PlotkinSS98,HummerG03}. 
In a system with many degrees of freedom on a complex energy landscape
and %%
obeying nontrivial steric restrictions, %%
finding a best reaction coordinate or even a good
reaction coordinate is a difficult task. Finding reaction
paths between metastable minima is an old problem, in which many approaches
have been developed to account for the underlying complex,
multi-dimensional potential energy
surface~\cite{CerjanCJ81,BellS84,ElberR87,WalesD93,WalesD01,KomatsuzakiT03,PrentissMC10}. 

An alternate approach, in the spirit of defining order parameters in
statistical and condensed matter physics,  is to consider the geometry of the product
and reactant in defining a reaction coordinate without reference to
the underlying potential energy landscape. The overlap function $q$ of
a spin-glass is an example of a geometrically-defined order
parameter~\cite{MezardM86}, for which the underlying Hamiltonian determines
behavior such as the temperature-dependence. We pursue such a geometric
approach in this paper. 

A transformation connecting unfolded states with the native folded
state can be considered as a reaction coordinate. A transformation can
also be used as a starting point for refinement, by examining commitment
probability or other reaction coordinate formalism. 

Several methods have been developed to find transformations between protein
conformational pairs without specific reference to a molecular
mechanical force field. These include 
coarse-grained elastic network models~\cite{KimMK02,KimMK02bj},
coarse-grained plastic network models~\cite{MaragakisP05}, iterative
cluster-normal mode analysis~\cite{SchuylerAD09},  restrained
interpolation (the Morph server)~\cite{KrebsWG00}, the FRODA
method~\cite{WellsS05}, and geometrical targeting (the geometrical
pathways (GP) server)~\cite{FarrellDW10}. 

In this paper we consider transformations between polymer conformation
pairs that would not be viable by a conjugate-gradient type or direct minimization 
approach, in that dead-ends would inevitably be encountered. We focus
specifically on how one might find geometrically optimal
transformations that account for polymer non-crossing constraints, which
would apply to knotted proteins for example. 

By a geometrically optimal transformation, we mean a transformation
in which every monomer in a polymer, as represented by the
$\alpha$-carbon backbone of a protein for example,  would travel the least distance
in 3-dimensional space in moving from conformation A to conformation
B. This is a variational problem, and the equations of motion, along
with the minimal transformation and the Euclidean distance covered, have been
worked out
previously~\cite{PlotkinSS07,MohazabAR08,MohazabAR08:bj,MohazabAR09}. Although
minimal transformations 
have been found for the backbones of secondary structures, and the non-crossing problem
has been treated~\cite{MohazabAR08:bj}, minimal transformations between unfolded and
folded states for full protein chain lengths have not been treated
before. %%

The minimal transformation
inevitably involves curvilinear motion if bond, angle, or
stereochemical constraints are
involved~\cite{GrosbergAY04,PlotkinSS07}. 
Such curvilinear transformations as a result of bond constraints were developed
in~\cite{PlotkinSS07,MohazabAR08,MohazabAR08:bj,MohazabAR09}.
If such constraints are
neglected, the minimal distance
corresponding to the minimal transformation reduces to the mean of
the root squared distance (MRSD), or the mean of the
straightline distances between pairs of atoms or monomers. This is not
the conventional RMSD. For any typical pair of conformations, the MRSD is always
less than the RMSD~\cite{MohazabAR08}. %%
Used as an alignment cost function, aligned configurations using MRSD are
globally different than those using RMSD~\cite{MohazabAR09}. 
The RMSD can be thought of as a least squares fit between the
coordinates defining the two
structures. Alternatively, it may also be thought of as the
straight-line Euclidean distance between two 
structures in a high-dimensional space of dimension $3 N$, where
$N$ is the number of atoms in the protein, or $C_\alpha$ atoms if the
protein is coarse-grained. Fast
algorithms have been constructed to align structures using
RMSD~\cite{KabschW76,KabschW78,KnellerGR91,FlowerDR99,CoutsiasEA04,CoutsiasEA05}.

If several intermediate states are known along the pathway of a
transformation between a pair of structures, then the RMSD may be
calculated consecutively for each successive pair. This notion of RMSD
as an order parameter goes back to reaction dynamics papers
from the early
1980's~\cite{CerjanCJ81,BellS84,ElberR87,WalesD93}, however in these
approaches the potential
energy governs the most likely reactive trajectories
taken by the system, and RMSD is simply accumulated through the
transition states. 

In the absence of a potential surface 
except for that corresponding to steric constraints, the incremental RMSD may be treated as
a cost function and %%
the corresponding transformation between two structures
found algorithmically~\cite{FarrellDW10}. 
However, the minimal transformation using RMSD (or $3ND$ Euclidean
distance) as a cost
function is different than the minimal transformation using $3D$
Euclidean distance (MRSD) as a cost function, and the RMSD-derived
transformation does not correspond to the
most straight-line trajectories.
The RMSD is not equivalent to the total amount of motion a protein or
polymer must undergo in transforming between structures, even in the
absence of steric constraints enforcing deviations from straight-line
motion.  Conversely, the transformation corresponding to the MRSD will be
curvilinear in the 3N-dimensional space.

In what follows, we develop a computational scheme for describing how
difficult it might be for different proteins to reach their folded
configuration. The essence is a calculation of how much "effort" the
protein chain must expend to avoid having to cross through itself as
it tries to realize its folded state.
This involves finding the different ways a polymer can 
uncross or ``untangle'' itself, and then calculating the corresponding
distance for each of the
untangling transformations. Since there are typically several avoided
crossings during a minimal folding transformation, finding the optimal
untangling strategy corresponds to finding the optimal 
combination of uncrossing operations with minimal total distance
cost. 

After quantifying such a procedure, we apply this to full length
protein backbone chains for several structural classes, including
$\alpha$-helical proteins, $\beta$-sheet proteins, $\alpha$-$\beta$
proteins, 2-state and 3-state folders, and knotted proteins. We
generate unfolded ensembles for each of the proteins investigated, and
calculate minimal distance transformations for each member of the
unfolded ensemble to fold. From this calculation, we obtain the mean minimal distance to fold
from the unfolded ensemble, for a given structural class. 
We look for differences in the mean minimal distance
between structural and kinetic classes, and compare these to
differences in other order parameters between the respective
classes. 
The extra non-crossing distance per residue
$\mathcal {D}_{nx}/N$ %%
turns out to be the most consistent
discriminator between different structural and kinetic classes of
proteins. We find the extra distance covered to avoid chain 
crossing is generally a small fraction ($\sim 1/10$) of the total
motion. We also investigate how the various order
parameters either correlate or are independent from each other. 

We then select three proteins, an $\alpha$-helical, a $\beta$-sheet,
and a knotted protein, to further dissect the taxonomy of their
minimal folding transformations. 
We construct what might be called ``multiple transformation
alignments'' that describe the various different ways each protein can
fold from an ensemble of unfolded conformations. 
We find that noncrossing motions of an N- or C-terminal leg are
generally obligatory for a knotted protein, and only incidental for an
$\alpha$ protein. A consensus minimal folding transformation is
constructed for each of the above-mentioned native folds, and rendered
schematically.  By investigating a ``pathway overlap'' order
parameter, we find that non-crossing constraints, as are prevalent in
$\beta$ proteins and pervasive in knotted proteins, explicitly induce
a pathway ``mechanism'' in protein folding, as defined by a common
sequence of events independent of the initial unfolded conformation. 
We finally discuss our results and conclude.

\section*{Methods}

\subsection*{Calculation of the transformation distance}

The value of the uncrossing or non-crossing distance, $\Dnx$, is calculated as follows:
The chain
transforms from conformation A to conformation B as a ghost chain, so
the chain is allowed to pass through itself.  The beads of the chain
follow  straight trajectories from initial to final positions. This
is an approximation to the actual Euclidean distance $\cal D$ of the
transformation, where straight line transformations of the beads are
generally preceded or proceeded by non-extensive local rotations to
preserve the link length connecting the beads as a rigid constraint~\cite{PlotkinSS07,MohazabAR08}. 
The instances of self-crossing along with their
times are recorded. The associated cost for these crossings is
computed retroactively, for example the distance cost for one arm of
the chain to circumnavigate another obstructing part is then added to
the ``ghost'' distance to compute the total distance.

The method for calculating the non-crossing distance $\Dnx$ has three major
components, evolution of the chain, crossing detection, and crossing
cost calculation. Each are described in the subsections below. 

\subsubsection*{Evolution of the chain}
\label {sec:evolution_of_the_chain}
As mentioned above, the condition of constant link length between
residues along the chain is relaxed, so that the non-extensive
rotations that would generally contribute to the distance traveled are
neglected here.  This approximation becomes progressively more
accurate for longer chains.  Thus ideal transformations only involves
pure straight-line motion. The approximate transformation is carried
out in a way to minimize deviations from the true transformation
($\cal D$), such that link lengths are kept as constant as possible,
given that all beads must follow straight-line motion. We thus only
allow deviations from constant link length when rotations would be
necessary to preserve it; this only occurs for a small fraction of the
total trajectory, typically either at the beginning or the end of the
transformation \cite{PlotkinSS07,MohazabAR08}.

\paragraph{A specific example}

As an example of the amount of distance neglected by this
approximation, consider the pair of configurations in
Figure~\ref{fig:method_delta}, where a chain of 10 residues that is
initially horizontal transforms to a vertical orientation as shown in
the figure. The distance neglecting rotations (our approximation) is
77.78, in reduced units of the link length, while the exact
calculation including rotations~\cite{PlotkinSS07,MohazabAR08} gives a distance of
78.56.

A few intermediate conformations are shown in the figure. In
particular note the link length change (and hence violation of
constant link length condition) in the fourth link for the gray
conformation (conformation F), resulting from our approximation. If
the link length is preserved, the transformation consists of local
rotations at the boundary points.

Also note that when transforming from cyan to magenta the first bead
moves less than $\delta$, because it reaches its final destination and
``sticks'' to the final point, and will not be moved subsequently.  A
movie of this transformation is provided as Movie 1 in the
Supplementary Material.

\paragraph{General method}
\label{sec:gen_method}

The algorithm to evolve the chain is as follows.  Straight-line paths
from the positions of the beads in the initial chain configuration to
the corresponding positions of the beads in the final configuration
are constructed. The bead furthest away from the destination, i.e. the
bead whose path is the longest line, is chosen.  Let this bead be
denoted by index $b$ where $0\le b \le N$. In the context of
figure~\ref{fig:method_delta}, this bead corresponds to bead number 9 ($b_9$).
The bead is then moved toward its destination by a small
pre-determined amount $\delta$, and the new position of bead $b$ is
recorded.  In this way the transformation is divided into say $M$
steps: $M=d_{max} / \delta$, where $d_{max}$ is the maximal
distance. Let $i$ be the step index $0 \le i \le M$.  If initially the
chain configuration was at step $i$ (e.g. $i=0$), the spatial position
of bead $b$ at step $i$ before the transformation $\delta$ is denoted
by ${\bf r}_{b,i}$, and after the transformation by ${\bf r}_{b,i+1}$.
The upper bound $\delta$ to capture the essence of the transformation
dynamics differs according to the complexity of the problem. 
To capture all of the instances of self-crossing,  a step size
$\delta$ of two percent of the link length sufficed for all 
cases.

The neighboring beads ($b+1$ and $b-1$) should also follow paths on
their corresponding straight-line trajectories.  Their new position on
their paths (${\bf r}_{b+1,i+1}$ and ${\bf r}_{b-1,i+1}$) are then
calculated based on the constant link length constraints. This new
position corresponds to moving the beads by $\delta_{b+1,i}$,
$\delta_{b-1,i}$ respectively. Once ${\bf r}_{b+1,i+1}$ and ${\bf
  r}_{b-1,i+1}$ are calculated, we proceed and calculate ${\bf
  r}_{b+2,i+1}$ and ${\bf r}_{b-2,i+1}$ until we reach the end points
of the chain. As an example consider figure~\ref{fig:method_delta},
going from the conformation B (Green) to the conformation C
(Yellow). First, bead number 9, which is the bead farthest away from
its final destination, is moved by $\delta$, then taking constant link
length constraints and straight line trajectories into account, the
new position of bead 8 is calculated and so on, until all the new bead
positions which correspond to the yellow conformation are calculated.

If somewhere during the propagation to the endpoints, a solution
cannot be constructed or no continuous solution exists,
i.e. $\lim_{\delta \to 0} ({\bf r}_{b+m,i+1} - {\bf r}_{b+m,i}) \ne
0$, then we set ${\bf r}_{b+m,i+1} = {\bf r}_{b+m,i}$. That is, the
bead will remain stationary for a period of time. \footnote{This
  in principle may result in a link length change for the
  corresponding link, and thus constraint violation, in our
  approximation. An exact algorithm involves local link rotation
  instead.} Consequently ${\bf r}_{b+n,i+1} = {\bf r}_{b+n,i}$ for all
beads with 
$n > m$ that have not yet reached their final destination. This is
because the new position of each bead is calculated by the
position of the bead next to it for any particular step $i$. 
The same recipe is applied when propagating incremental motions
$\delta_{b,i+1}$ along the other direction of the chain (going from
$b-n$ to $b-n-1$) as well.  
When a given bead that has been held stationary becomes 
the farthese bead away from its final position, it is then moved
again. 
I.e. stationary beads can move
again at a later time during the transformation if they become the
furthest beads away from the final conformational state.  Such a
scenario does not occur in the context of the simple example of
figure~\ref{fig:method_delta}, however in Movie 2 in the Supplementary
Material, a transformation is given for a full protein that involves
such a process.  During the course of such a transformation the viewer
will notice that several beads on the chain (in the upper right in the
movie) remain stationary for a
part of the transformation. For these beads no continuous solution for
the motion exists, i.e. as $\delta \rightarrow 0$ the beads in
question cannot move without violating the constant link length
constraint. 
At a later time during the transformation, when the beads in the given segment are
farthest from the final folded conformation, the beads resume motion.

Once the positions of all the beads in step $i+1$ are calculated, the
same procedure is repeated for step $i+2$ and so on, until the chain
reaches the final configuration. If the position of a given bead $b$
at step $i$ is such that $|{\bf r}_{b,i} - {\bf R}_{b}| < \delta$, where
${\bf R}_{b}$ is the spatial position of bead $b$ in the final conformation,
then ${\bf r}_{b,i+1}$ is set to ${\bf R}_{b}$. In other words we discretely
snap the bead to the final position if it is closer than the step size
$\delta$.  In the context of figure~\ref{fig:method_delta}, this
corresponds to going from conformation D (Cyan) to conformation E
(Magenta). Bead 0 ($b_0$) is snapped to the final conformation. Once a bead
reaches its destination it locks there and will never move again. See
conformation F (gray) in figure~\ref{fig:method_delta}.

Figure~\ref{fig:link-deviation} shows a histogram of the mean link
length over the course of a transformation,
for 200 transformations between random
initial structures generated by self avoiding random walks (SAW), and one
pre-specified SAW. The length of the random chains was 9 links. The
chains were aligned by minimizing MRSD before the transformation took
place~\cite{MohazabAR08,MohazabAR08:bj,MohazabAR09}, where MRSD
stands for the mean root squared distance and is defined by ${1 \over
  N} \sum_{n=1}^N \sqrt{(\bfr_{A_{n}} - \bfr_{B_{n}})^2} = {1 \over
  N} \sum_{n=1}^N | \bfr_{A_{n}} - \bfr_{B_{n}} |$.
Deviations from the  full unperturbed 
link length are modest: the ensemble-averaged mean link length is 96\%
of the initial link length.

\subsubsection*{Crossing Detection}
\label{sec:crossing_detection}
As stated earlier, during the transformation the chain is initially
treated as a ghost chain, and so is allowed to cross itself. To keep
track of the crossing instances of the chain, a crossing matrix
$\mathbb{X}$ is updated at all time steps during the
transformation. If the chain has $N$ beads and $N-1$ links, we can
define an $(N-1) \times (N-1)$ matrix $\XX$ that contains the crossing
properties of a 2D projection of the strand, in analogy with
topological analysis of knots.  The element $\XX_{ij}$ is nonzero if
link $i$ is crossing link $j$ in the 2D projection at that
instant. Without loss of generality we can assume that the projection
is onto the XY plane, as in Figure~\ref{fig:3linkchainX}. We
illustrate the independence of our method on projection plane
explicitly for a crossing event in cold-shock protein (1CSP) 
in the Supplementary Material. We use the
XY plane projection throughout this paper.\footnote{We use the
  crossings in the projected image as a book-keeping device to detect
  real 3D crossings. A real crossing event is characterized by a sudden
  change in the over-under nature of a crossing on a projected
  plane. Since for any 3D crossing, the change of nature of the
  over-under order of crossing links is present in any arbitrary
  projection of choice, keeping track of a single projection is enough
  to detect 3D crossings (A concrete
  illustration of the independence of crossing detection on the
  projection plane is given in the Supplementary Material).  Of course
  a given projection plane may not 
  be the optimal projection plane for a given crossing, however if the
  time step is small enough any projection plane will be sufficient to
  detect a crossing.}

We parametrize the chain uniformly and continuously in the direction
of ascending link number by a parameter $s$ with range $0 \le s \le
N$. So for example the middle of the second link is specified by $s =
3/2$.  If the projection of link $i$ is crossing the projection of
link $j$, then $|\mathbb{X}_{ij}|$ is the value of $s$ at the crossing
point of link $i$ and $|\XX_{ji}|$ is the value of $s$ at the crossing
point of link $j$. If link $i$ is over $j$ (i.e. the corresponding
point of the cross on link $i$ has a higher $z$ value than the
corresponding point of the cross on link $j$) then $\XX_{ij} > 0$,
otherwise $\XX_{ij}<0$. Thus after the sign operation,
$sign(\mathbb{X})$ is an anti-symmetric matrix.

A simple illustrative example of the value of $\XX$ for the 3-link chain
in figures~\ref{fig:3linkchainX}a and  \ref{fig:3linkchainX}b is

\begin{subequations}
  \label{eq:eqcross}
  \begin{eqnarray}
    \label{eq:eqcross1}
    \XX \left(t_o \right) &=& 
    \begin{bmatrix}
      0   &  0 & -0.29\\
      0   &  0 &   0\\
      +2.82&  0& 0
    \end{bmatrix}\\
    \label{eq:eqcross2}
    \XX \left(t_o + \delta \right)  &=& \begin{bmatrix}
      0   &  0 & +0.29\\
      0   &  0 &   0\\
      -2.82&  0& 0
    \end{bmatrix} 
  \end{eqnarray}
\end{subequations}

The fact that $\XX_{13}$ is negative at time $t_o$ indicates that at
that instant, link 1 is under link 3 in 3D space, above the
corresponding point on the plane on which the projections of the links
have crossed (green circle in figure \ref{fig:3linkchainX}).

At each step during the transformation of the chain, the matrix $\XX$
is updated. A true crossing event is detected by looking at $\XX$ for
two consecutive conformations. A crossing event occurs when any
non-zero element in the matrix $\XX$ discontinuously changes sign
without passing through zero. Once $\XX_{ij}$ changes sign, $\XX_{ji}$
must change sign as well.  If the chain navigates through a series of
conformations that changes the crossing sense and thus the sign of
$\XX_{ij}$, but does not pass through itself in the process, the
matrix elements $\XX_{ij}$ will not change sign discontinuously but
will have values of zero at intermediate times before changing sign.

Movie 3 in the Supplementary Material shows the result of applying
crossing detection. In the movie of the transformation, whenever an
instance of self-crossing is detected, the transformation is halted
and the image is rotated to make the location of the crossings easier
to visualize.

\subsubsection*{Crossing Cost calculation}
\label{sec:crossing_cost_calculation}
Even in the simplest case of crossing, there are multiple ways for the
real chain to have avoided crossing itself.  The extra distance that
the chain must have traveled during the transformation to respect the
fact that the chain cannot pass through itself is called the
``non-crossing'' distance $\Dnx$.  If the chain were a ghost
chain which could pass through itself, the corresponding distance for
the whole transformation would be the MRSD, along with relatively
small modifications that account for the presence of a conserved link
length.  Accounting for non-crossing always introduces extra distance
to be traveled.

As the chain is transforming from conformation A to conformation B as
a ghost chain according to the procedure discussed
above, a number of self-crossing incidents occur.
Figure~\ref{fig:simple_untangle} shows a continuous but topologically
equivalent version of the crossing event shown in
figure~\ref{fig:3linkchainX} (b).  Even for this simple case, there
are multiple ways for the transformation to have avoided the crossing
event, each with a different cost.

Furthermore, later crossings can determine the best course of action
for the previous crossings.  Figure \ref{fig:retrospect_matters}
illustrates how non-crossing distances are non-additive, so that one
must look at the whole collection of crossing events.  Therefore to
find the optimum way to ``untangle'' the chain (reverse the sense of
the crossings), one must look at all possible uncrossing
transformations, in retrospect. The recipe we follow is to evolve the
chain as a ghost chain and write down all the incidents of
self-crossings that happen during the transformation. Then looking at
the global transformation, we find the best untangling movement that
the chain could have taken.

To compute the extra cost introduced by non-crossing constraints we
proceed as follows: We construct a matrix that we call the cumulative
crossing matrix $\YY$. $\YY_{ij}$ is non-zero if link $i$ has truly
(in 3D) crossed link $j$, at any time during the transformation.  This
matrix is thus conceptually different than the matrix $\XX$, which
holds only for one instant (one conformation) and which can have
crossings in the 2D projection which are not true crossings during the
transformation.  The values of the elements of $\YY$ are calculated in
the same way that the values are calculated for $\XX$. The sign also
depends on whether the link was crossed from over to under or from
under to over, so that a given projection plane is still assumed. The
order in which the crossing have happened are kept track of in another
matrix $\YY_O$. The coordinates of all the beads at the instant of a
given crossing are also recorded. For example, if during the
transformation of a chain, two crossing have happened, then two sets
of coordinates for intermediate states are also stored. We describe a
simple concrete example to illustrate the general method next.

\paragraph{A Concrete Example}

Figure~\ref{fig:concrete_eg} shows a simple transformation of a 7-link
chain. During the transformation the chain crosses itself in two
instances. The first instance of self-crossing is between link 5 and
link 7. The second instance is when link 2 crosses link 4. The
location of the crossing along the chain is also recorded: i.e. if we
assume that the chain is parametrized by $s = 0$ to $N$, then at the
instant of the first crossing (link 5 and link 7) $s = 4.4$ (link 5)
and $s = 6.9$ (link 7). The second crossing occurs at $s=1.3$ (link 2)
and $s=3.8$ (link 4). The full coordinates of all beads are also
known: we separately record the full coordinates of all beads at each
instant of crossing. The information that indicates which
links have crossed and their over-under structure can be aggregated
into the cumulative crossing matrix $\YY$. For the example in
figure~\ref{fig:concrete_eg}, the cumulative matrix (up to a minus sign
indicating what plane the crossing events have been projected on) is
\[
\YY = \begin{bmatrix}
  
  0&   0&   0&   0&   0&   0&   0\\
  0&   0&   0&   1.3& 0&   0&   0\\
  0&   0&   0&   0&   0&   0&   0\\
  0&  -3.8& 0&   0&   0&   0&   0\\
  0&   0&   0&   0&   0&   0&  -4.4\\
  0&   0&   0&   0&   0&   0&   0\\
  0&   0&   0&   0&   6.9& 0&   0
\end{bmatrix} \: .
\]
$\YY$ tells us, during the whole process of transformation, which
links have truly crossed one another and what the relative over-under
structure has been at the time of crossing. For example, by glancing at
the matrix we can see that two links 5,7 and 2,4 have crossed one
another.  We also know from the sign of the elements in $\YY$ that
both links 2 and 7 were underneath links 4 and 5 just prior to their
respective crossings in the reference frame of the projection.
Two links will cross each other at most once during a
transformation. If one link, e.g. link $i$, crosses several others
during the transformation, elements $i,j$, $i,k$ etc... along with their
transposes will be nonzero.

The order of crossings can be represented in a similar fashion as a
sparse matrix.
\[
\YY_O = \begin{bmatrix}
  
  0&   0&   0&   0&   0&   0&   0\\
  0&   0&   0&   2&   0&   0&   0\\
  0&   0&   0&   0&   0&   0&   0\\
  0&   2&   0&   0&   0&   0&   0\\
  0&   0&   0&   0&   0&   0&   1\\
  0&   0&   0&   0&   0&   0&   0\\
  0&   0&   0&   0&   1&   0&   0
\end{bmatrix}.
\]

Analyzing the structure of the crossings is similar to analyzing the
structure of a knot, wherein one studies a knot's 2D projections,
noting the crossings and their over/under nature based on a given
directional parameterization of the curve
\cite{AdamsCCKnot,NechaevSKStatKnot,WiegelFW86}.  One difference here
is that we are not dealing with true closed-curve knots (in the mathematical
sense), as a knot is a representation of S1 in S3. Here we treat open curves.

\subsubsection*{Crossing substructures}
\label{sec:substructures}

By studying the crossing structure of open-ended pseudo-knots in the most
general sense, one can identify a number of sub-structures that recur
in crossing transformations. Any act of reversing the nature of all
the crossings of the polymer can be cast within the framework of some
ordered combination of reversing the crossings of these substructures.

We identify three sub-structures: Leg, Loop, and Elbow. 
\paragraph{Leg}
\label{sec:leg}

Given any self-crossing point of a chain, a leg is defined from that
crossing point to the end of the chain.  Therefore for each
self-crossing point two legs can identified as the shortest distance
along the chain from that crossing point to each end---see
figure~\ref{fig:leg_example}.  A single leg structure is shown in
figure~\ref{fig:all_object}(a).

\paragraph{Loop}
\label{sec:loop}
As stated earlier, when traveling along the polymer one arrives at
each crossing twice. If the two instances of a single crossing are
encountered consecutively while traveling along the polymer, and no
intermediate crossing occurs, then the substructure that was traced in between
is a loop. See Fig~\ref{fig:all_object}(b).

\paragraph{Elbow}
\label{sec:elbow}
If two consecutive crossings have same over-under sense, then they
form an elbow; see Fig~\ref{fig:all_object}(c). Note that the same
two consecutive crossing instances will occur in reverse order on the
second visit of the crossings: these form a dual of the elbow. By
convention the segment with longer arc-length between the two
consecutive crossings is defined as the elbow. This would be the
horseshoe shaped strand in figure \ref{fig:all_object}(c).

\subsubsection*{Reversing the crossing nature}
The goal of this formalism is to assist in finding a series of
movements that will result in reversing the over-under nature of all
the crossings, with the least amount of movement required by the
polymer. So at this point we introduce basic movements that that will
reverse the nature of the crossings for the above substructures.

\paragraph{Using leg movement}
\label{sec:legmove}
A transformation that reverses the over/under nature of a leg involves
the motion of all the beads constituting the leg. Each bead must move
to the location of the crossing (the ``root'' of the leg), and then
move back to its original location~\cite{MohazabAR08:bj}. The canonical
leg movement is shown schematically in figure \ref{fig:leg_movement}.

We can reverse the nature of all the crossings that have occurred on a
leg, if more than one crossing occurs, through a leg movement (see
figure~\ref{fig:leg_movement_extra_cross}). The move is topologically
equivalent to the movement of the free end of the leg along the leg up
to the desired crossing, and then moving all the way back to the
original position while reversing the nature of the crossing on
the way back.

\paragraph{Loop twist and loop collapse}
\label{sec:looptwist}
Reversing the crossing of a loop substructure can be achieved by a
move that is topologically equivalent to a twist, see
figure~\ref{fig:loop_twist_all} (a). This type of move is called a
Reidemeister type I move in knot theory. However the optimal motion is
generally not a twist or rotation in 3-dimensional space (3D). Figure
\ref{fig:loop_twist_all}(b) shows a move which is topologically
equivalent to a twist in 3D, but costs a smaller distance, by simply
moving the residues inside the loop in straight lines to their final
positions, resulting in a ``pinching'' motion to close the loop and
re-open it. From now on we refer to the optimal motion simply as loop
twist, because it is topologically equivalent, but we keep in mind
that the actual optimal physical move, and the distance calculated
from it, is different.

\paragraph{Elbow moves}
\label{sec:elbowmove}
Reversing the crossings of an elbow substructure can be done by moving
the elbow segment in the motion depicted in
figure~\ref{fig:elbow_move}: Each segment moves in a straight line to
its corresponding closest point on the obstruction chain, and then it moves in a
straight line to its final position.

\subsubsection*{Operator Notation}
\label{sec:operatornotatoin}
The transformations for leg movement, elbow, and twist can be expressed
very naturally in terms of operator notation, where in order to
untangle the chain the various operators are applied on the chain
until the nature of all the self-crossing are reversed.

If we uniquely identify each instance of self-crossing by a number,
then a topological loop twist at crossing $i$ can be represented by
the operator $R(i)$ ({\it R} for Reidemeister). An elbow move, for the
elbow defined by crossings $i$ and $i+1$, can be represented as
$E(i,i+1)$. As discussed above, for each self-crossing, two legs can
be identified corresponding to the two termini of the chain. This was
exemplified in figure~\ref{fig:leg_example}, by the red and blue
legs. Since we choose a direction of parametrization for the chain, we
refer to the two leg movements as the ``start leg'' movement and the
``end leg'' movement, and for a generic crossing $i$ we denote them as
$L_N(i)$ and $L_C(i)$ respectively.

The operators that we defined above are left acting (similar to matrix
multiplication). So a loop twist at crossing $i$ followed by
an elbow move at crossings $j$ and $j+1$ is represented by $E(j,j+1)
R(i)$.

\paragraph{Example}
\label{sec:operatornotatoin_example}
Figure~\ref{fig:operator_example} shows sample configurations before
and after untangling. The direction of parametrization is from the red
terminus to the cyan terminus. It can be seen that there are several
ways to untangle the chain. One example would be $R(3)L_C(2)R(1)$,
which consists of a twist of the green loop, followed by the cyan leg
movement, followed by a twist of the blue loop.  Another path of
untangling would be $E(2,3)L_N(1)$,
which is movement of the red leg followed by the magenta elbow move.

For the two above transformations, the order of operations can be
swapped, i.e. they are commutative, and the resulting distance for
each of the transformations will be the same. That is
$\mathcal{D}[E(2,3)L_N(1)] = \mathcal{D}[L_N(1)E(2,3)]$. However,
$E(2,3)L_N(1)$ is a more efficient transformation than
$R(3)L_C(2)R(1)$, i.e. 
$\mathcal{D}[E(2,3)L_N(1)]< \mathcal{D}[R(3)L_C(2)R(1)]$.

Other transformation moves are not commutative in the algorithm, for
example in Figure~\ref{fig:operator_example},  $L_N(1)R(3)R(2)$ is not
allowed, since $R(2)$ will only act on loops defined by two instances
of a crossing that are encountered consecutively in traversing the
polymer, i.e. no intermediate crossings can occur. Therefore even if
crossing 2 happens kinetically before crossing 3 during the ghost transformation,
only transformation $L_N(1) R(2) R(3)$ is allowed in the algorithm.

\subsubsection*{Minimal uncrossing cost}
\label{sec:minimal_untangling_cost}
For each operator in the above formalism, a transformation distance/cost can
be calculated. Hence the optimal untangling strategy is finding the
optimal set of operator applications with minimal total cost. This
solution amounts to a search in the tree of all possible
transformations, as illustrated in Figure~\ref{fig:tree_of_poss}. The
optimal application of operators can be computed by applying a
version of the depth-first tree search algorithm.

According to the algorithm, from any given conformation there are several moves
that can be performed, each having a cost associated with the
move. The pseudo-code for the search algorithm can be written as
follows:  %%
\begin{verbatim}
 procedure find_min_cost (moves_so_far=None, cost_so_far=0,\
                         min_total_cost=Infinity):
    optim_moves = NULL_MOVE
    if cost_so_far > min_total_cost:
        return [Infinity, optim_moves]
    endif

    for move in available_moves(moves_so_far):
        [temp_cost, temp_optim_moves] = find_min_cost (moves_so_far + move,\
                                                  cost_so_far + cost(move),\
                                                  min_total_cost)
        if temp_cost < min_total_cost:
            min_total_cost = temp_cost
            optim_moves = move + temp_optim_moves
        endif
    endfor
   
    return [min_total_cost,optim_moves]
endprocedure
\end{verbatim}
The values to the right side of the equality sign in the arguments of
the procedure are the default values that the procedure starts
with. The procedure is called recursively, and returns both the set
of optimal uncrossing moves (for a given crossing matrix corresponding
to a starting and final conformation),
and the distance corresponding to that
set of optimal uncrossing moves. 

The algorithm visits all branches of the tree of possible uncrossing
operations until it reaches the end. However it is smart enough to
terminate the search along the branch if the cost of operations
exceeds that of a solution already found. See
figure~\ref{fig:tree_of_poss} for an illustraion of the depth-first search tree
algorithm.
The above procedure was implemented using both the GNU Octave
programing language and C++. To optimize speed by eliminating
redundant moves, only one permutation was considered when operators
commuted.

\subsection*{Generating unfolded ensembles}
\label{sec:generate_unfold}
To generate transformations between unfolded and folded conformations,
we adopt an off-lattice coarse grained $C_\alpha$ model
\cite{ClementiC00:jmb,SheaJE00}, and generated an unfolded structural
ensemble from the native structure as follows. For a native structure
with $N$ links, we define three data sets:
\begin{itemize}
\item{The set of $C_\alpha$ residue indices $i$, for which $i=1
    \cdots N$ }
\item{The set of native link angles $\theta_j$ between three
    consecutive $C_\alpha$ atoms, for which $j=2 \cdots N-1$}
\item{The set of native dihedral angles $\phi_k$ between four
    consecutive $C_\alpha$ atoms, i.e. the angle between the
    planes defined $C_\alpha$ atoms ($k-1$, $k$, and $k+1$), and ($k$,
    $k+1$, and $k+2$). The index $k$ runs from $k=2\cdots N-2$.}
\end{itemize}

The distribution of C$_\alpha$-C$_\alpha$ distances in PDB
structures is sharply peaked around 3.76\AA~ ($\sigma=0.09$\AA). 
In practice we took the first C$_\alpha$-C$_\alpha$
distance from the N-terminus 
as representative, and used that number for the equilibrium link length
for all C$_\alpha$-C$_\alpha$ distances in the protein. 

To generate an unfolded ensemble, we start by selecting at random a
$C_\alpha$ atom $n$ ($2 \le n \le N-1$)
in the native conformation, and we then
perform rotations that change the angle centered at that randomly
chosen residue $n$, $\theta_n$, and that change the dihedral %%
defined by rotations about the bond $n$-($n$+1), $\phi_n$. 
If $n=N-1$ only the angle is changed.  The new angle and dihedral are
selected at random from the Boltzmann distributions as described
below. 
After each rotation, $\theta_n \to \theta_n^{new}$ and $\phi_n
\to \phi_n^{new}$. Changing these 
angles rotates the entire rest of the chain, i.e. all the
beads $i$ with $i > n$ %%
are rotated to a new
position. This recipe corresponds to an extension of the pivot
algorithm \cite{LalM69,MadrasN88}. 

However, we additionally require that the values of each angle
and dihedral that are present in the native structure,
$\theta_n^{Nat}$ and $\phi_n^{Nat}$, are more
likely to be observed. We implement this criterion in the following way.
The new angle $\theta_n$ is chosen from a probability distribution
proportional to $\exp{\left(-\beta E \left(\theta_n\right) \right)}$,
where $\beta E(\theta_n)$ is computed
from: 
\begin{equation}
  \label{eq:EDeltatheta}
  \beta E(\theta_n) =  k_\theta \left(\theta_n -
    \theta_n^{Nat}\right)^2 \: ,
\end{equation}
where we have set $k_\theta = 20$. 
Similarly for the dihedral $\phi_n$, the probability distribution function is
proportional to $\exp{\left(-\beta E \left(\phi_n\right)\right)}$, where
$\beta E \left(\phi_n\right)$ is computed from
\begin{equation}
  \label{eq:EDeltaphi}
  \beta E \left(\phi_n\right) =  k_{\phi1} [1 + \cos{( \phi_n -
    \phi_n^{Nat})}] +  k_{\phi3} [1 + \cos{(3  (\phi_n -
    \phi_n^{Nat}))}]   \: ,
\end{equation}
where $k_{\phi1} = 1$, and 
$k_{\phi 3} = 0.5$. %%
The fact that the $k_\phi$s are much smaller
than $k_\theta$ means that for a given temperature, dihedral angles
are more uniformly distributed than bond angles. If 
all $k_\theta$ and $k_\phi$ are 
set to zero, then all states are equally accessible and the algorithm
reduces to the pivot algorithm, i.e. a generator for unbiased,
self-avoiding random walks. If
all $k_\theta$ and $k_\phi$ are 
set to $\infty$, then chain behaves as a rigid object and
does not deviate from its native state.

Each pivot operation results in a new structure that must be checked
so that it has no steric overlap with itself, i.e. the chain must be
self-avoiding. If the new chain conformation has steric overlap, then
the attempted move is discarded, and a new residue is selected at
random for a pivot operation.  

In practice, we defined steric overlap
by first finding an approximate contact or cut-off distance for the coarse-grained
model. The contact distance was taken to be the smaller of either the
minimum $C_\alpha$-$C_\alpha$ distance between those residues in
native contact (where two residues are defined to be in native contact if any of
their heavy atoms are within 4.9\AA), or the $C_\alpha$-$C_\alpha$
distance between the first two consecutive residues.
For SH3 for example the minimum $C_\alpha$ distance in native contacts
is $4.21$\AA~and the first link length is $3.77$\AA, so
for SH3 all non-neighbor beads must be further than $3.77$\AA~for a
pivot move to be accepted.  Future refinements of the acceptance
criteria can involve the use of either the mean C$_\alpha$-C$_\alpha$
distance or other criteria more accurantly representing the steric
excluded volume of residue side chains.

In our recipe, to generate a single unfolded structure we start with
the native structure and implement $\mathcal{N}$ {\it successful}
pivot moves,
where $\mathcal{N}$ is related to the number of residues $N$ by
$\mathcal{N} = \ln(0.01)/{\ln[0.99 (N-2)/(N-1)]}$. 

For the next unfolded structure we start again from the native
structure  and pivot $\mathcal{N}$ successful times,
following the above recipe. Note that $\mathcal{N}$ successful pivots
does not generally affect all beads of the chain. In the most likely
scenario some beads are chosen several times and some beads are not
chosen at all, according to a Poisson distribution. This particular
choice of $\mathcal{N}$ means that for polymers with $N<101$ where
${N-2 \over N-1} < 0.99$, the chance that any given link is not
pivoted at all during the $\mathcal{N}$ pivot operations is $0.01$. On
the other hand for longer polymers where ${N-2 \over N-1} > 0.99$, the
probability that any particular segment of the protein with the length
$0.01$ of the total length, has $0.01$ chance of not having any of its
beads pivoted. For any $N$ however, the shear number of pivot moves
generally ensures a large RMSD between the native and generated
unfolded structures. 

Each unfolded structure generally retains small amounts of
native-like secondary and tertiary structure, due to the native biases
in angle and dihedral distrubutions.  For example, for SH3 the number
of successful pivot moves was 162 and the mean fraction of native contacts in the
generated unfolded ensemble was $0.06$.

\subsection*{Protein dataset}

The 45 proteins used in this study are given in
Table~\ref{tab:proteins_used}. 
When divided into kinetic classes, they consist of 25 2-state
folders, 13 non-knotted
3-state folders, and 7 knotted proteins not used in the kinetic analysis. 
Structurally there are  11 all $\alpha$-helix proteins, 14 all $\beta$-sheet
proteins, 13 $\alpha$-$\beta$ proteins, and 5 knotted proteins.  %%
These proteins were selected randomly from the datasets in
references~\cite{IvankovDN03,GromihaMM06}, 
where kinetic rate data was available to categorize the proteins
into 2-state or 3-state folders.
Our dataset contains 27
out of the 52 proteins
in~\cite{IvankovDN03}, and 38 out
of the 72 proteins in~\cite{GromihaMM06}.
The datasets in~\cite{IvankovDN03,GromihaMM06}
do not include knotted
proteins however; the Knotted proteins were taken from several
additional sources, including references~\cite{MallamAL07}
(1NS5),~\cite{MallamAL06} (1MXI),~\cite{KingNP10} (3MLG),  
~\cite{vanRoonAMM08}  (2K0A),~\cite{BolingerD10} (2EFV), 
and the protein knot server KNOTS~\cite{KolesovG07} 
(1O6D, 2HA8).  Aside from the Stevadore knot  in~\cite{BolingerD10} 
we did not consider pseudo-knots more complex than the $3_1$  trefoil. 

Several of these proteins ($\alpha$-amylase inhibitor 2AIT and MerP
mercuric ion binding protein 2HQI) have disulfide bonds present in the
native structure. These constraints are not used in the current
analysis. The folding pathways we obtain may be thought of as relevant
to the initial folding event before disulfide bonds are formed, or
for a protein of equivalent topology but sequence lacking the disulfide
bond. Lack of preservation of disulfide bonds is a shortcoming of the
present algorithm; development of more accurate computational algorithms for
unfolded ensemble generation are a topic of future work.

Several of the proteins also have ligands present in the crystal
or NMR structures. These include 1A6N and 1HRC (heme ligands),  1RA9
(Nicotinamide adenine), 1GXT (sulfate), 1MXI (iodide ion), 2K0A (3
Zn ions), 2EFV (phosphate ion). Since we have removed energetics in
general from our analysis of geometrical pathways, these ligands and any effect they
may have on the folding pathway due to protein-ligand interactions
are not included here. In the folding kinetics analysis of 
references~\cite{IvankovDN03,GromihaMM06}, 
they are generally not present either, e.g. the folding rate for 1A6N is
actually that for apomyoglobin~\cite{CavagneroS99}. 

\subsubsection*{Structural alignment properties of our protein dataset}
To categorize proteins as two- or three-state, we have chosen
proteins with folding rate data available. This dataset has somewhat
different structural alignment statistics than that for a non-redundant (NR)
database, e.g.~\cite{Thiruv2005nh3d}. 
The TM-score based alignment of Zhang and Skolnick~\cite{ZhangY05}
can be used to obtain structural alignment statistics. Their
method resolves the problems of outlier and length-dependent
artifacts of 
RMSD-based alignments. Distributions of TM-score for both the above
NR database, our dataset, and the datasets in references~\cite{IvankovDN03,GromihaMM06},
which non-knotted
proteins in our dataset were taken, are given in the Supplementary
Material along with statistical analysis of the
distributions. The bulk of our proteins
(98\%) have TM-scores consistent with the NR database of 
Thiruv {\it et. al}
(see Figure~S2 in the Supporting Information), however our dataset and those of~\cite{IvankovDN03,GromihaMM06} 
contain a small number of structural homologs not
present in the NR dataset, which
are tabulated in the Supplementary Material. 
We do not suspect that this small number of homologs will
significantly modify the conclusions derived from statistical analysis of
our dataset, however expansion and refinement to find the most
relevant dataset is a topic for future work. 

\subsection*{Calculating distance metrics for the unfolded ensemble}
\label{sec:details_of_method}
To obtain minimal transformations between unfolded and native
structures for a given protein, the $C_\alpha$ backbone was extracted
from the PDB native structure, and 200 coarse-grained unfolded
structures were generated using the methods described above. The unfolded
structures were then aligned using RMSD and the average (residual)
RMSD was calculated. The unfolded structures were then aligned by
minimizing MRSD, and the residual MRSD was calculated. Then
conformations were further coarse-grained (smoothed) by sampling every
other bead, hence reducing the total number of beads. By the above
further-coarse graining which is in the spirit of
the initial steps of Koniaris-Muthukumar-Taylor
reduction~\cite{KoniarisK91,KoniarisK91-JCP,TaylorWR00,VirnauP05},  
we eliminate all instances of potential
self-crossing in which the loop size or elbow size is smaller than
three links. Each structure was then transformed to the folded state
by the algorithm discussed earlier in Methods. 
The self-crossing instances, along with the coordinates of all the
beads, were recorded as well.
Appropriate data structures were formed and relevant crossing
substructures (leg, elbow, and loop) were detected. With topological data structures
at hand, the minimal uncrossing cost was found, through the 
depth-first search in the tree of possible uncrossing operations that
was described above.
Finally, the minimal uncrossing cost, $\Dnx$, and the total distance,
$\mathcal{D}$ are calculated for each unfolded conformation. These
differ from one unfolded conformation to
the other; the ensemble average is recorded and used below. 
The ensemble average of MRSD and RMSD are also calculated from the 200
unfolded structures that were generated.\\

\subsubsection*{Importance of non-crossing}
\label{importance_of_non_crossing}
We define the importance of non-crossing (INX) as the ratio of the extra
untangling movement caused by non-crossing constraints, divided by the
distance when no such constraints exists, i.e. if the chain behaved as a
ghost chain. Mathematically this ratio is defined as $INX = \Dnx /
\left(  \mbox{\sc mrsd} \times N \right)$

\subsubsection*{Other metrics}
Other metrics investigated include absolute
contact order ACO \cite{Plaxco98}, relative contact order RCO
\cite{Plaxco98}, long-range order LRO \cite{GromihaMM01},
and chain length N\cite{GutinAM96:prl,GalzitskayaOV01}.  

Following~\cite{GromihaMM01}, we define Long-range Order (LRO) as:
\begin{equation}
  \label{eq:LRO_define}
  LRO = \sum_{i<j} n_{ij} / N \,\,  \mbox { where }
  n_{ij} = 
  \begin{cases}
    1 &  \mbox{if }|i - j | > 12 \\
    0 & \mbox{otherwise}
  \end{cases}
\end{equation}
where i and j are the sequence indices for two residues for which the
$C_\alpha - C_\alpha$ distance is $\le 8$ \AA~in the native
structure. 

Likewise we define Relative Contact Order (RCO) following \cite{Plaxco98}:
\begin{equation}
  \label{eq:RCO_define}
  RCO = {1 \over L \times N} \sum_{i<j}^N \Delta L_{ij}, 
\end{equation}
where $N$ is the total number of contacts between nonhydrogen atoms in
the protein that are within 6 \AA~in the native structure,  $L$ is the
number of residues, 
and $\Delta L_{ij}$ is the sequence separation between contacts in units
of the number of residues.

Similarly, Absolute Contact order (ACO)~\cite{Plaxco98} is defined to be:
\begin{equation}
  \label{eq:ACO_define}
  ACO = {1 \over  N} \sum_{i<j}^N \Delta L_{ij} = RCO \times L 
\end{equation}

\section*{Results}
Proteins were classified by several criteria: 
\begin{itemize}
\item {2-state vs. 3-state folders}
\item {$\alpha$-helix dominated, vs $\beta$-sheet dominated, vs mixed.}
\item {knotted vs unknotted proteins}
\end{itemize}

\noindent Several questions are answered for each group of proteins:
\begin{itemize}
\item {What fraction of the total transformation distance is due to
    non-crossing constraints?}
\item {How do the different order parameters distinguish between the
    different classes of proteins?}
\item {How do the different order parameters correlate with each other?}
\end{itemize}

\subsubsection*{Order paramaters discriminate protein classes}
In table~\ref{tab:order_parameters_for_various_classes}, we compare
the unfolded ensemble-average of several metrics between different classes of
proteins, and perform a p-value analysis based on the Welch
t-test. The null hypothesis states that the two samples being compared
come from normal distributions that have the same means but possibly
different variances. Metrics compared in
Table~\ref{tab:order_parameters_for_various_classes} are INX, LRO,
RCO, ACO, MRDS, RMSD, $\Dnx$, $\Dnx/N$, $\mathcal{D}$, $\mathcal{D}/N$
and $N$.

The most obvious check of the general method outlined in the present
paper is to compare the non-crossing distance $\Dnx$ between knotted
and unknotted proteins. Here we see that knotted proteins traverse
about $3.5\times$ the distance as unknotted proteins in avoiding
crossings, so that the two classes of proteins are different by this
metric. The same conclusion holds for knotted {\it vs.} unknotted
proteins if we use $\Dnx/N$, $\mathcal{D}$, $\mathcal{D}/N$, or INX. Of all
metrics, the statistical significance is highest when comparing
$\mathcal{D}/N$, which is important because the knotted proteins
considered here tend to be significantly longer than the unknotted
proteins, so that chain length $N$ distinguishes the two
classes. Dividing by $N$ partially normalizes the chain-length dependence of
$\Dist$, however $\Dist/N$
still correlates remarkably strongly with $N$ when compared for all proteins
($r=0.824$ see Supplementary Material, Table 8).

It was somewhat unusual that MRSD and RMSD distinguished knotted proteins from
unknotted proteins better than $\Dist$ (or $\Dnx$), which accounts for
non-crossing. All other quantities, including INX, ACO, and
RCO distinguish knotted from unknotted proteins. The only quantity
that fails is LRO. 

The importance of noncrossing $INX$, measuring the ratio of the
uncrossing distance $\Dnx$ to the ghost-chain distance $N\times MRSD$,
was largest for knotted proteins, followed by $\beta$ proteins, with
$\alpha$ proteins having the smallest $INX$. Mixed proteins had an
average INX value in between that for $\alpha$ and $\beta$ proteins. 

In distinguishing all-$\alpha$ and all-$\beta$ proteins, 
we find that LRO and RCO are by far the best
discriminants. Interestingly, INX and $\Dnx/N$ also discrimate these two classes
comparably or better than ACO does. $\Dnx$ is marginal, while all
other metrics fail. 

All metrics except for $N$ and $\Dist$ are able to discriminate
$\alpha$ from mixed $\alpha$-$\beta$ proteins, with LRO performing the
best by far. Interestingly, none of the above metrics can distinguish $\beta$ proteins
from mixed $\alpha$-$\beta$ proteins.

It is sensible that energetic considerations would be the dominant
distinguishing mechanism between two- and three state folders.
Intermediates are typically stabilized energetically.  We can
nevertheless investigate whether any geometrical quantities
discriminates the two classes. Indeed LRO and RCO fail, as does
INX. This supports the notion that intermediates are not governed by
``topological traps'' that are undone by uncrossing motion, but rather
are energetically driven.  ACO performs marginally. Three-state
folders tend to be longer than 2-state folders, so that $N$
distinguishes them and in fact provides the strongest
discriminant, consistent with previous
results~\cite{GalzitskayaOV03}. Interestingly RMSD, MRSD, and $\Dist$
perform comparably 
to $N$. However these measures also correlate strongly with N
(see Supplementary Information Table 8).  $\Dist/N$, $\Dnx$ and
$\Dnx/N$ also perform well, but still correlate with N, albeit more
weakly than the above metrics.

Figure~\ref{fig:cluster}A shows a scatter plot of all proteins as a
function of $\Dnx/N$ {\it vs.} and LRO. Knotted and unknotted proteins
are indicated, as are $\alpha$, $\beta$, and mixed $\alpha$-$\beta$
proteins. Two and three state proteins are indicated as triangles and
squares respectively.  From the figure, it is easy to visualize how LRO
provides a successful discriminant between $\alpha/\beta$ and
$\alpha/$(mixed) proteins, but is unsuccessful in discriminating
$\beta/$(mixed), knotted and unknotted, and two and three state
folders.  It is also clear from the figure how $\Dnx/N$ discriminates
knotted from unknotted proteins. One can also see distribution
overlap, but nevertheless successful discrimination between $\alpha$
and $\beta$ and $\alpha$ and mixed proteins.

Figure~\ref{fig:cluster}B shows a scatter plot of all proteins as a
function of $\Dnx$ {\it vs.} $N$, using the same rendering scheme for
protein classes as in Figure~\ref{fig:cluster}A. From the figure, one
can see how the metrics correlate with each other, and how they both
discriminate knotted from unknotted proteins and 2-state from 3-state
proteins. Moreover one can see how despite the significant correlation
between $\Dnx$ and $N$, $\Dnx$ can discriminate $\alpha$ proteins from
either $\beta$ proteins or mixed $\alpha$/$\beta$ proteins, while N
cannot. 

As a control study for the above metrics, we took random selections
of half of the proteins, to see if random partitioning of the proteins
into two classes resulted in any of the metrics distinguishing the two
sets with statistical significance. No metric in this study had
significance: the p-values ranged from about 0.32 to 0.94.

Figure~\ref{fig:class_distinguish} shows a plot of the statistical
significance for all the metrics in
Table~\ref{tab:order_parameters_for_various_classes} to distinguish
various pairs of proteins classes: 2-state from 3-state proteins, $\alpha$ from
$\beta$, $\alpha$ from mixed $\alpha/\beta$, $\beta$ from mixed, and
knotted from unknotted. 
We can define the most consistent discriminator between protein classes as
that metric that is statistically significant for the most classes,
and for those classes has the highest statistical significance. By
this criterion $\Dnx/N$ is the most consistent discriminator between
the general structural and kinetic classes considered here.

Interestingly, in all cases, the extra distance
introduced by non-crossing constraints is a very small fraction (less
than 13\% ) of the MRSD, 
which represents the ghost distance neglecting
non-crossing.  This was not an obvious result, but
it was encouraging evidence for the reason simple order-parameters
that neglect an explicit accounting of non-crossing have been so
successful historically~\cite{Onuchic96,Plaxco98,Baker2000:Nature,NymeyerH00:pnas,BestRB05,DingF05jmb,ChoSS06}.

\subsubsection*{Scaling laws for pathway distances across domains and whole proteins} 

Larger proteins will typically have larger MRSD. A protein of twice
the chain length need not have twice the MRSD however; we plot the
unfolded ensemble averaged MRSD of the proteins in
our dataset as a function of $N$ in Figure~\ref{figscaling}A. The plot shows
sub-extensive scaling for the straight-line path distance per residue:
$MRSD \sim N^{0.65}$. On the other hand, the 
non-crossing distance per residue, $\Dnx/N$, shows superextensive
scaling:  $\Dnx/N \sim N^{1.33}$, indicating that non-crossing induced
entanglement becomes progressively more important even on a
per-residue basis for longer proteins, and likely polymers in
general.  In fact, the steeper slope of $\Dnx/N$ indicates a crossover
such that when $N$ is larger than about $3600$, chain non-crossing
dominates the motion of the minimal folding pathway. 
It is noteworthy that the scatter in the log-log scaling plot of
Figure~\ref{figscaling}A  is much larger for $\Dnx/N$ than for MRSD, illustrating the larger
dispersion of $\Dnx/N$ for proteins of the same length but
different native topology.

The above analysis can be applied to domains within a single
protein, to test how autonomous their folding mechanisms are as 
compared to separate proteins.  
Run on our dataset, the program DDomain~\cite{ZhouH09ddomain}
only finds multiple domains in 
methyltransferase domain of human tar (HIV-1) RNA binding protein
(PDB 2HA8)~\cite{WuH2HA8}, between residues 20-88 and 89-178 (residue 20
is the first resolved residue in the crystal structure).
The domain finding program DHcL~\cite{KoczykG08}
also finds domains in this protein between residues 20-83 and
84-178. DHcL also finds domains in several other proteins, some
generally accepted as single domain, however one of these proteins 
is clearly a repeat protein containing a 36 residue helix-turn-helix motif: 
tumor suppressor P16INK4A (PDB 2A5E)~\cite{ByeonIJL1998}.
For this protein, DHcL finds domains between the 1st and 2nd, and
2nd and 3rd repeating units. We manually
added a domain boundary between the 3rd and 4th repeating units to
yield 4 domains containing residues 1-36, 37-72, 73-108, and
109-144. The domains of 2HA8 and 2A5E are illustrated in Figure~\ref{figscaling}C.

Using the above domain structures for 2HA8 and 2A5E,  we analyze the
scaling of MRSD with chain length N in Figure~\ref{figscaling}B.
In these plots the
individual domains are considered as separate proteins, then
combined together if the domains are contiguous, e.g. for 2A5E proteins consisting of domains 1,
domains 1 and
2 together, 1, 2, and 3 together, all domains together, and all
contiguous combinations therein are
examined. This yields the
same scaling law for both proteins: $MRSD \sim N^{0.76}$, which has a
larger power law than the scaling between proteins above. 
Chain connectivity constraints apparently induce cross-talk between 
domains even for MRSD. 
Likewise, the
scaling law for noncrossing distance per residue is  $\Dnx/N \sim N^{2.51}$, 
indicating significant polymer chain interference between domain
folding. The individual domains of multidomain proteins apparently
show less severe chain constraints than single domain proteins of the
same size.

\subsubsection*{Quantifying minimal folding pathways}

The minimum folding pathway gives the most direct way that an unfolded
protein conformation can transform by reconfiguration to the native
structure. However, different configurations in the unfolded ensemble
transform by different sequences of events, for example one unfolded
conformation may require a leg uncrossing move, followed by a
Reidemeister move elsewhere on the chain, followed by an uncrossing
move of the opposite leg, while another unfolded conformation may
require only a single leg uncrossing move. 

The sequence of moves can be represented as a color-coded bar plot, as
shown in
Figures~\ref{figtransformsalpha}-\ref{figtransformsknot}. 
In these figures, the sequence of moves is taken from right
to left, and the width of the bar indicates the non-crossing distance
undertaken by that move. 
A scale bar is given underneath each figure indicating a distance of
100 in units of the link length. 
Red bars indicate
moves corresponding to the N-terminal leg ($\LN$) of the protein, while green
bars indicate moves corresponding to the C-terminal leg ($\LC$). Blue bars
indicate Reidemeister ``pinch and twist'' moves, while cyan bars
indicate elbow uncrossing moves.

The typical sequence of moves varies depending on the protein.  
Figure~\ref{figtransformsalpha} shows the uncrossing
transformations of the all-$\alpha$ protein acyl-coenzyme A binding
protein (PDB id 2ABD~\cite{AndersenKV93}, see Figure~\ref{fig:3protsrender}A). 
Panels A and B depict the same set of
transformations, but in A they are sorted from largest to smallest
values of $\LN$ uncrossing, and in B they are sorted from
largest to smallest values of $\LC$ uncrossing. The leg moves
in each panel are aligned so that the left end of the bars
corresponding to the moves being sorted are all lined up.
Some transformations partway down in panel A do not require an $\LN$ move; these are then
ordered from largest to smallest $\LC$ move.  The converse is applied
in panel B. Some moves do not require either leg move; these are
sorted in decreasing order of the total distance of Reidemeister loop
twist moves. Finally, some transformations require only elbow moves;
these are sorted from largest to smallest total uncrossing distance.

Figure~\ref{figtransformsbeta} shows the
noncrossing transformations for
the Src homology 3 (SH3) domain of phosphatidylinositol 3-kinase (PI3K),
a largely-$\beta$ protein (about 23\% helix, including 3 short
$3_{10}$ helical turns; PDB id 1PKS~\cite{KoyamaS93}, see
Figure~\ref{fig:3protsrender}B), sorted analogously to 
Figure~\ref{figtransformsalpha}. 
Figure~\ref{figtransformsknot} shows the
uncrossing transformations involved in the minimal folding of the
designed knotted protein 2ouf-knot (PDB id 3MLG~\cite{KingNP10},
Figure~\ref{fig:3protsrender}C).  

Interestingly, for the all-$\alpha$ protein 2ABD, $\approx 12\%$ of the
172 transformations considered did not require any uncrossing moves,
and proceed directly from the unfolded to the folded conformation. These
transformations are not
shown in Figure~\ref{figtransformsalpha}. For the $\beta$ protein and
knotted protein, every transformation that we considered (195 for 1PKS
and 90 for 3MLG) required at least one uncrossing move.

As a specific example, the top-most move in Figure~\ref{figtransformsknot}
panel B consists of a
C-leg move (green) covering $\approx 90\%$ of the non-crossing
distance, followed by N-leg move (red) covering $\approx 7 \%$ of
the distance, then a short elbow move (cyan), a short
Reidemeister loop move (blue), another short elbow move (cyan), and finally a short Reidemeister move
(blue). In some cases the elbow and loop moves commute if they involve
different parts of the chain, but generally they do not. For this
reason we have not made any attempt to cluster loop and elbow moves,
rather we have just represented them in the order they occur. On the other hand, consecutive
leg moves commute and can be taken in either order.

In Figures~\ref{figtransformsalpha}-\ref{figtransformsknot}, 
one can see that significantly more motion is involved in the leg
uncrossing moves than for other types of move. 
The total distance covered by leg moves is 82\% for 3MLG, 69\% for
1PKS, and 49\% for 2ABD. For 3MLG, the total leg move distance is comprised of 44\% 
$\LN$ moves, and 38\% $\LC$ moves. For 1PKS, leg move distance is
comprised of 18\% $\LN$ moves, and 51\% $\LC$ moves. For 2ABD,
distance for the leg moves is roughly symmetric with 
26\% $\LN$ and 23\% $\LC$.

One difference that can be seen for the all-$\alpha$ protein compared
to the $\beta$ and knotted proteins  is in the persistence of the leg
motion. For 2ABD, only 24\% of the transformations require
$\LN$ moves and only 30 \% of the transformations require $\LC$
moves. On the other hand the persistence of leg moves is greater in
the $\beta$ protein and greatest in the knotted protein. For 1PKS,
$\LN$ and $\LC$ moves persist in 74\% and 66 \% of the
transformations respectively. In 3MLG, 
$\LN$ and $\LC$ moves persist in 92 \% and 41 \% of the
transformations respectively. 

Inspection of the transformations for the $\beta$ protein 1PKS in
panels A and B of Figure~\ref{figtransformsbeta} reveals that
uncrossing moves generally cover larger distance than in the $\alpha$
protein 2ABD (the mean uncrossing distance for is 136 for 1PKS {\it vs.}
77.5 for 2ABD). We also notice that in contrast to the leg
uncrossing moves in 2ABD, both $\LN$ and $\LC$ moves are often
required (44 \% of the transformations require both $\LN$ and
$\LC$ moves, compared to 5 \% for 2ABD). 
The asymmetry of the protein is manifested in the asymmetry of the leg
move distance: the $\LN$ moves are generally shorter than the $\LC$
moves, covering about 1/4 of the total leg move
distance. As mentioned above, $\LC$ moves comprise  
about 51 \% of the total distance for the 195 transformations in
\ref{figtransformsbeta}, while $\LN$ moves only comprise about
18 \% of the distance on average.
Both $\LN$ and $\LC$ moves are persistent as mentioned above. A leg move
of either type is present in 95 \% of the transformations.

Inspection of the transformations in Figure~\ref{figtransformsknot} reveals
that every transformations requires either an $\LN$ or $\LC$ move. 
This is sensible for a knotted protein, and is in contrast to the
transformations for the $\alpha$ protein 
2ABD, where many moves do not require any leg uncrossing at all and consist of
only short Reidemeister loop and elbow moves. In this sense the
diversity of folding routes~\cite{PlotkinSS00:pnas,PlotkinSS02:Tjcp} for the knotted protein 3MLG is the
smallest of the proteins considered here, and illustrates the concept
that topological constraints induce a pathway-like aspect to the folding mechanism.
The N-terminal $\LN$ leg move is
the most persistently required uncrossing move, present in about
92\% of the transformations. This is generally the terminal
end of the protein that we found was involved in forming the pseudo-trefoil
knot. Sometimes however, the C-terminal end is involved in forming the
knot, though
this move is less persistent and is present in only 41\% of
the transformations. However when
an $\LC$ move is undertaken, the distance traversed is significantly
greater, as shown in  Panel B of Figure~\ref{figtransformsknot}. 
This asymmetry is a 
consequence of the asymmetry already present in the native structure
of the protein.

\subsubsection*{Consensus minimal folding pathways}

From the transformations described in
Figures~\ref{figtransformsalpha}-\ref{figtransformsknot}, we see that
there are a multitude of different transformations that can fold each
protein. The pathways for the $\alpha$ protein 2ABD are more diverse
than those for the $\beta$ or knotted proteins.  From
the ensemble of transformations for each protein, we can average the
amount of motion for each uncrossing move to obtain a quantity
representing the consensus or most representative minimal folding
pathway for that protein.  This takes the form of the histograms in
Figure~\ref{fig:consensus_moves}, with the x-axes representing the
order of uncrossing/untangling events, right to left, and the y-axes
representing the average amount of motion in each type of move.

The ensemble of untangling transformations can be divided into three different
classes: transformations in which leg $L_N$ is the largest move, transformations in
which leg $L_C$ is the largest move, and transformations in which an
elbow E or loop R (for Reidemeister type I) are the largest
moves. Moreover, if $L_N$ and
$L_C$ moves occur consecutively they can be commuted, so without loss
of generality we take the $L_N$ move as occurring before the $L_C$
move in the x-axes of Figure~\ref{fig:consensus_moves}. The leg
moves, if they occur first, are then followed by either elbow (E) and/or
loop (R) moves, of which there may be several. 
In general, the leg moves may both occur before the collection of loop and elbow
moves, after them, or may bracket the elbow and loop moves 
(e.g. 2nd bar in Figure~\ref{figtransformsknot}b). 
By the construction of our approximate algorithm, if two $\LN$ moves
were encountered during a trajectory (they were encountered only a few
times during the course of our studies), they would be aggregated into one
$\LN$ move involving the larger of the two motions, in order to remove
any possible redundancy of motion. Hence no more than one $\LN$ or
$\LC$ move is obtained for all transformations.  
We found
that three pairs of elbow and loop moves was sufficient to desribe
about 93 \% of all transformations (see the x-axes of
Figure~\ref{fig:consensus_moves}). 
In summary, the sequence $L_N$, $L_C$,
$R$, $E$, $R$, $E$, $R$, $E$, $L_N$, $L_C$ (read from left to right)
characterized almost all transformations, and so was adopted as a general scheme.
Any exceptions simply had more small elbow and loop moves that were of
minor consequence; for these transformations we simply accumulated 
the extra elbow and loop moves into the most appropriate $R$ or $E$
move. The general recipe for rendering loops R in
Figure~\ref{fig:consensus_moves} is as follows: if one  R move is
encountered (regardless of where), each half is placed first and last
(third) in the general scheme. If two R moves are encountered, they
are placed first and last, and if three R moves are encountered, they
are simply partitioned in the order they occured. For four or more R
moves, the middle $N-2$ are accumulated into the middle slot in the
general scheme. The same recipe is applied to elbow moves E. 
As a specific example, the first bar in
Figure~\ref{figtransformsknot}b consists of $L_C$, $L_N$, $E_1$, $R_1$, $E_2$, $R_2$,
which after permutation of the first two leg moves falls into the general
scheme above as $L_N$, $L_C$, $R_1$, $E_1$, $0$, $0$, $R_2$, $E_2$, $0$, 
$0$. 
The bottom-most transformation in
Figure~\ref{figtransformsknot}B consists of $R_1$, $R_2$, $R_3$, $E_1$, $E_2$, $E_3$, $L_N$, 
which becomes
$0$, $0$, $R_1$, $E_1$, $R_2$,
$E_2$, $R_3$, $E_3$, $L_N$, $0$ in the general scheme.

Figure~\ref{fig:consensus_moves} shows histograms of the minimal
folding mechanisms, obtained from the above-described procedure. Note
again there are 3 classes of transformation, one where $\LN$ is the
largest move, one where $\LC$ is the largest move, and one where either
loop R or elbow E is the largest move. 
Each uncrossing element of the transformation,
C-leg, N-leg, Reidemeister loop, or elbow, contributes to the height of the
corresponding bar, which represents the average over transformations
in that class. 
The percentage of transformations that fall into each class is given
in the legend to panels A-C of Figure~\ref{fig:consensus_moves}. 

Most of the transformations ($71.5\%$)  for
the $\alpha$-protein 2ABD fall into the class with a dominant loop or
elbow move, which itself tends to cover less uncrossing distance than
either C- or N-leg uncrossing (ordinates of Panels A-C
Figure~\ref{fig:consensus_moves}). 
This is a signature of a diverse range of folding
pathways: minimal folding pathways need not involve obligatory leg
uncrossing constraints. In this sense, the $\beta$ protein 1PKS is
more has a more constrained folding mechanism than the $\alpha$
protein; there is a significantly larger percentage of
transformations for which a leg transformation $\LC$ or $\LN$
dominates, though the mean distances undertaken when a leg move does
dominate are comparable for $\LC$ and even larger for the $\alpha$
protein for $\LN$. 

The knotted protein 3MLG has the most constrained minimal folding
pathway. A leg move from either end dominates for 91\% of the
cases. Even for the transformations where loop or elbow moves
dominate, there is still relatively significant $\LN$ motion. The dominant pathways for
knotting 3MLG involve leg crossing from either N or C terminus. When
the C terminus is involved in the minimal transformation, the motion
can be significant (Figure~\ref{fig:consensus_moves}B).

Among all transformations of a given protein, a transformation can be
found that is closest to the average transformation for one of the
three classes in Figure~\ref{fig:consensus_moves}.  This consensus
transformation has a sequence of moves that when mapped to the scheme
in Figure~\ref{fig:consensus_moves}, has minimal deviations from the
averages shown there. Further, we can find the transformation that has
minimal deviation to any of the three classes in
Figure~\ref{fig:consensus_moves}. For the knotted protein 3MLG, the
best fit transformation is to the class with $\LN$-dominated moves, for
the $\alpha$ protein 2ABD, the best fit transformation is to the class
with miscellaneous-dominated moves, and the $\beta$ protein 1PKS, the
best fit transformation is the class with $\LC$-dominated moves.  For
the $\alpha$, $\beta$, and knotted proteins, these are the
transformations denoted by a short arrows to the left of the
transformation in panels A and B of
Figures~\ref{figtransformsalpha},~\ref{figtransformsbeta},
and~\ref{figtransformsknot} respectively. For the $\alpha$, $\beta$,
and knotted proteins, the transformations are illustrated
schematically in Figures~\ref{fig:2abd_move},~\ref{fig:1pks_move},
and~\ref{fig:3mlg_move} respectively.

Inspection of the most
representative transformation for the all-$\alpha$ protein 2ABD shown
in  Figure~\ref{fig:2abd_move} indicates 
that the transformation requires remarkably little motion:
it contains a negligible leg motion followed by a loop uncrossing of
modest distance, followed by a short elbow move that is also
inconsequential: in shorthand $E[9] R[20] \LN[1]$, where the
numbers in brackets indicate the cost of the moves in units where the
link length is unity.
In constructing a schematic of the representative transformation in
Figure~\ref{fig:2abd_move}, 
we ignore the smaller leg and elbow moves and illustrate the loop move
roughly to scale. Although additional crossing points appear from the
perspective of the figure, the remainder of the transformation
involves simple straight-line motion.

Figure~\ref{fig:1pks_move} shows the most representative folding
transformation for the $\beta$ protein 1PKS. The sequence of events
constructed from the most representative minimal transformation,
$E[18] R[24] \LC[48]$, consists of a dominant leg move depicted
in steps 4 and 5 of the transformation, followed by shorter loop and
elbow Reidemeister moves that are neglected in the
schematic. Loops and crossing points appear
from the perspective of the figure, however the remainder of the
transformation involves simple straight-line motion.

Figure~\ref{fig:3mlg_move} shows the most representative folding
transformation for the knotted protein 3MLG. The sequence of events
constructed from the minimal transformation, $R[21] R[18]
\LN[125]$ in the above notation, consists of a
dominant leg move depicted in steps 4 and 5 of the transformation, and
two relatively short loop moves that are neglected in the schematic as
inconsequential.
Loops appear from the perspective of the figure, and the
crossing points appear to shift in position, however the remainder of the
transformation involves simple straight-line motion.

\subsubsection*{Topological constraints induce folding pathways} 

From Figures~\ref{figtransformsalpha}-\ref{figtransformsknot}, one can
see that topological non-crossing constraints can induce pathway-like
folding mechanisms, particularly for knotted proteins, and in part for
$\beta$-sheet proteins as well. The locality of interactions in conjunction
with simple tertiary arrangement of helices in the 
$\alpha$-helical protein profoundly affects the nature of the
transformations that fold the protein, such that the distribution of
minimal folding pathways is diverse. Conversely, the knotted protein,
although largely helical, has non-trivial tertiary arrangement, which
is manifested in the persistence of a leg crossing move in the
minimal folding pathway. In this way, a folding ``mechanism'' is induced
by the geometry of the native structure. 

We can quantify this notion by calculating the similarity between
minimal folding pathways. To this end we note that, for example, the
transformation that is 
6 bars from the bottom in Figure~\ref{figtransformsknot}b, which contains
an $\LN$ move followed by 2 short loops and an elbow, 
should not fundamentally be very different than the transformation 10 from
the bottom in that figure, which contains a loop and 2 short elbows followed
by a larger $\LN$ move. 
In general we treat the commonality of the moves as relevant to the
overlap rather than the specific number of residues involved, or the
order of the moves that arises from the depth-first tree search
algorithm.

Thus for each transformation pair we define two sequence overlap
vectors in the following way. Overlaying the residues involved in
moves for each transformation along the primary sequence on top of
each other as in Figure~\ref{fig:sequence_move_match}, we count as
unity those moves of the
same type that overlap in sequence for both transformations, otherwise
a given move is assigned a value of zero. So for
example in Figure~\ref{fig:sequence_move_match} the result is two vectors of binary
numbers, one with 4 elements for transformation $\alpha$ and one with
5 elements for transformation $\beta$, based on the
overlap of moves of the same type. That is, the first vector is 
$\vec{\Delta}^\alpha = (1,1,0,1)$ and the 2nd vector is $\vec{\Delta}^\beta =
(1,0,1,0,1)$. To find the pathway overlap, we also record the noncrossing
distances of the various transformations which here would be two
vectors of the form 
$\vec{\Dist}^\alpha =
(\Dist_{\LN}^\alpha,\Dist_{R_1}^\alpha,\Dist_{R_2}^\alpha,\Dist_{\LC}^\alpha
)^{\intercal}$, and 
$\vec{\Dist}^\beta =
(\Dist_{\LN}^\beta,\Dist_{R_1}^\beta,\Dist_{R_2}^\beta,\Dist_{E_1}^\beta,\Dist_{\LC}^\beta
)^{\intercal}$.
Square matrices 
$\mathbb{\Delta}$ are constructed for $\alpha$ and $\beta$,
where each row is identical and equal 
to the vector $\vec{\Delta}$.
This matrix then operates on 
$\vec{\Dist}$ to make a new vector that has distances for the
elements that are nonzero in $\vec{\Delta}$, and is the same length
for both $\alpha$ and $\beta$. In the above example shown in
Figure~\ref{fig:sequence_move_match}, 
$\mathbb{\Delta}^\alpha  \vec{\Dist}^\alpha =
(\Dist_{\LN}^\alpha,\Dist_{R_1}^\alpha,\Dist_{\LC}^\alpha
)^{\intercal}$ and 
$\mathbb{\Delta}^\beta \vec{\Dist}^\beta =
(\Dist_{\LN}^\beta,\Dist_{R_2}^\beta,\Dist_{\LC}^\beta
)^{\intercal}$. These vectors are then multiplied through the inner
product, and divided by the norms of $\vec{\Dist}^\alpha$ and 
$\vec{\Dist}^\beta$ to obtain the overlap $Q^{\alpha\beta}$. In
the above example,  
$Q^{\alpha\beta} = (\Dist_{\LN}^\alpha \Dist_{\LN}^\beta +
\Dist_{R_1}^\alpha \Dist_{R_2}^\beta + \Dist_{\LC}^\alpha
\Dist_{\LC}^\beta )/\sqrt{ \sum_i \left(\Dist^\alpha\right)^2_i 
  \sum_j (\Dist^\beta )^2_j} $. 
In general, the formula for the overlap is given by 
\begin{equation}
  Q^{\alpha \beta} = \frac{ ( \mathbb{\Delta}^\alpha
    \vec{\Dist}^\alpha )
    \cdot ( \mathbb{\Delta}^\beta \vec{\Dist}^\beta )}{
    \sqrt{  (\vec{\Dist}^\alpha \cdot  \vec{\Dist}^\alpha )
      (\vec{\Dist}^\beta \cdot  \vec{\Dist}^\beta ) }
  } \: .
  \label{eq:Qab}
\end{equation}

When $\alpha=\beta$, $Q^{\alpha \beta} = 1$.  In the above example,
$Q^{\alpha \beta} < 1$ even if all loops were aligned, because there
is no elbow move in transformation $\alpha$.  If two transformations
have an identical set of moves, $Q^{\alpha \beta} =1$ if all the moves
have at least partial overlap with a move of the same type
in primary sequence. 
If a loop move in transformation $\beta$ overlaps two
loop moves in transformation $\alpha$, it is assigned to the loop with
larger overlap in primary sequence.  For the first two transformations
in Figure~\ref{figtransformsknot}A, $\Qab = 0.988$, 
and for the first two
transformations in Figure~\ref{figtransformsknot}B, $\Qab=0.999$. 
On the other hand for the first and last transformations in
Figure~\ref{figtransformsknot}B, $\Qab=0.033$.  

Figure~\ref{fig:Qdist} shows the distributions of overlaps $\Qab$ between
all pairs of transformations indicated in
Figures~\ref{figtransformsalpha}-\ref{figtransformsknot}, for the
three proteins shown in
Figure~\ref{fig:3protsrender}. 
The distributions show a transition from multiple diverse minimal
folding pathways for the $\alpha$ protein, to the emergence of a
dominant minimal folding pathway for the knotted protein. 
The mean overlap
$Q$ between transformations can be obtained by averaging $\Qab$ in
Equation~\eqref{eq:Qab} over all pairs of transformations:
$Q = \sum_{\alpha < \beta} \Qab / \left(N \left(N-1 \right)/2
\right)$. Mean overlaps for each protein are given in the caption to
Figure~\ref{fig:Qdist}. 
This illustrates that
topological constraints induce mechanistic pathways in protein
folding. We elaborate on this in the Discussion section.

\section*{Discussion}

The Euclidean distance between points can be generalized
mathematically to find the distance between polymer curves; this can
be used to find the minimal folding transformation of a protein.
Here, we have developed a method for calculating approximately minimal
transformations between unfolded and folded states that account for
polymer non-crossing constraints.  The extra motion due to
non-crossing constraints was calculated retroactively for all crossing
events of a ghost chain transformation involving straightline motion
of all beads on a coarse-grained model chain containing every other
C$\alpha$ atom, from an 
ensemble of unfolded conformations, to the folded structure as defined
by the coordinates in the protein databank archive. 
The distances undertaken by the uncrossing
events correspond to straight-line motions of all the beads from the
conformation before the crossing event, over and around the
constraining polymer, and back to the essentially identical polymer
conformation immediately after the crossing event.  Given a set of
chain crossing events, the various ways of undoing the crossings are
explored using a depth-first tree search algorithm, and the
transformation of least distance is recorded as the minimal
transformation.

We found that knotted proteins quite sensibly must undergo more
noncrossing motion to fold than unknotted proteins. We also find a
similar conclusion for transformations between all-$\beta$ and
all-$\alpha$ proteins; all-$\alpha$ proteins generally undergo very
little uncrossing motion during folding. In fact the %%
uncrossing
distance, $\Dnx$, averaged over the unfolded ensemble, can be
used as a discrimination measure between 
various structural and kinetic classes of proteins. 
Comparing several metrics arising from this work with several common
metrics in the literature such as RMSD, absolute contact order ACO,
and long range order LRO, we found that the most reliable
discriminator between structural classes, as well as between two- and
three-state proteins, was %%
$\Dnx$ per residue.
(later paragraph moved here:)
Knotted proteins, as
compared to unknotted proteins, are the most distinguishable class of
those we investigated, in that all metrics we investigated except for LRO
significantly differentiated the knotted from unknotted proteins. 
The differentiation between
structural or kinetic classes of proteins as studied here is a separate issue from the
question of which order
parameters may best correlate with folding rates %%
\cite{PlaxcoKW00:biochem,GromihaMM01,IvankovDN03,OztopB04,IstominAY07};
this latter question is an interesting topic 
of future research. Differentiating %%
native-structure based
order parameters that provide good correlates of folding kinetics is a
complicated issue, in that diffferent structural classes may correlate
better or worse with a given order parameter~\cite{IstominAY07}. 

Non-crossing distance per residue $\Dnx/N$ increases more rapidly with
chain length than the mean straight line distance between residue
pairs (MRSD). Considering proteins as separate domains indicates a
crossover at long chain length, about $N=3600$. Considering 
proteins built up by adding successive domains, specifically for 
two representative multi-domain proteins in
our dataset (2HA8 and 2A5E), indicates a crossover to
entanglement-dominated folding mechanisms at shorter lengths- about
$N=400$. This crossover point may indicate a regime where energetics
begins to play a role in order to fold domains independently, and avoid
progressively more significant polymer disentanglement in order to fold. 
Even for knotted proteins, the motion involved in avoiding
non-crossing constraints is only about 13\% of the total ghost chain
motion undertaken had the noncrossing constraints been neglected. This
was not an obvious result, to these authors at least. In contrast to
melts of long polymers, chain non-crossing and the resultant
entanglement does not appear to be a significant factor in protein
folding, at least for the structures and ensembles we have studied
here. 
It is tempting to conclude from this that chain non-crossing
constraints play a minor role in determining folding mechanisms. 
It is nevertheless an empirical fact that knotted proteins fold
significantly slower than unknotted proteins~\cite{MallamAL08,KingNP10}. 
As well, raw percentages of total motion do not take into account the
difficulty in certain types of special polymer movement, in particular
when the entropy of folding routes is tightly
constrained~\cite{PlotkinSS00:pnas,PlotkinSS02:Tjcp,PlotkinSS02:quartrev2,ChavezLL04,NorcrossTS06,FergusonA09}. 
The small percentage of non-crossing motion may offer some explanation
however, as to why simple order
parameters, such as absolute contact order, that do not explicitly account for noncrossing in
characterizing folding mechanisms have historically been so successful
in predicting kinetics. 

The non-crossing distance was calculated here for a chain of zero thickness,
so that non-crossing is decoupled from steric constraints. Finite
volume steric effects would likely enhance the importance of
non-crossing constraints, since the volume of phase space where chains
are non-overlapping is reduced, and thus chain motions must be further
altered to respect these additional
constraints~\cite{BanavarJR05}. Steric constraints may significantly
alter the shape of reactive trajectories, and slow kinetics by enforcing entropic
bottlenecks. Such constraints may become particularly important for
collapsed or semi-collapsed proteins, and knotted proteins where they
restrict stereochemically-allowed folding pathways.  These effects may
in principle be treated by extending the present formalism to include
non-zero chain thickness, and by extending the minimal folding pathway
to the partition function of pathways, with each pathway having weight
proportional to the exponent of the distance~\cite{PlotkinSS07}.
Such a treatment is an interesting and important topic of future work.

One potential issue in the construction of the algorithm used here is
that the approximated minimal transformation is generally not equivalent to a
kinetically realizable transformation. In the depth-first tree search
algorithm illustrated in Figure~\ref{fig:tree_of_poss}, the set of
crossing points defines a set of uncrossing moves that may be permuted,
or combined for example through a compound leg movement as in
Figure~\ref{fig:leg_movement_extra_cross}. However the 
kinetic sequence of crossing events, in particular those significantly
separated in ``time'' along the minimal transformation, may not be permutable or
combinable physically, at least not without modifying the distance
travelled.\footnote{As a
  hypothetical example, suppose at time $t_1$ a crossing event occurs
  between residue $a$ which is 10 residues in from the N-terminus, and
  residue $b$ somewhere else along the chain. Then at time $t_2$, the
  next crossing event involves a residue $c$ that is 20 residues in from
  the N-terminus, and residue $d$ somewhere along the chain. To avoid
  redundant motion, the minimal transformation is only taken to involve
  a leg motion between the residues from $c$ to the N-terminus, about
  point $d$; this is assumed to encompass the motion in the
  first leg transformation, even though the crossing events occured at
  different times.}
Hence the transformations are treated here as
approximations to the true minimal transformations that respect
non-crossing.

The algorithm as described above may underrepresent the amount of
motion involved in noncrossing by allowing kinetically separated moves
to be commutable. On the other hand, the motion assumed in the
algorithm to be undertaken by a
crossing event contains abrupt changes in the direction of the
velocity (corners) at the time of the uncrossing event, and so 
is larger than the true minimal distance. %%
These errors cancel at least
in part. It is an interesting topic of future research to develop an improved
algorithm that computes minimal transformations, perhaps using these
approximate transformations as a starting point for further
optimization or modification.

The mathematical construction of minimal folding transformations can
elucidate folding pathways. To this end we have dissected the morphology of protein structure
formation for several different native structures. We found that the
folding transformations of knotted proteins, and to a
lesser extent $\beta$ proteins, are dominated by persistent leg uncrossing
moves, while $\alpha$ proteins have diverse folding pathways dominated simply
by loop uncrossing. 

A pathway overlap function can then be defined, the structure of which
is fundamentally different for $\alpha$ proteins than for knotted
proteins. While the overlap function supports the notion of a diverse
collection of folding pathways for the $\alpha$ protein, the overlap
function for the knotted protein indicates that 
topological polymer constraints can induce “mechanism” into how a
protein folds, i.e. these constraints induce a dominant sequence of events in the folding
pathway. This effect is observed to some extent in the $\beta$ protein
we investigated, but is most pronounced for knotted proteins.

Other approaches have been made previously to quantify topological
frustration, and construct folding pathways that minimize such
frustration. Norcross and Yeates~\cite{NorcrossTS06} have extended the
earlier analysis of Connolly {\it et. al}~\cite{ConnollyM80}, to 
show that %%
edges 
between consecutaive $C_\alpha$ atoms in the coarse-grained primary
sequence %%
can be surrounded by a ring of other $C_\alpha$ atoms
consisting of the
vertices of tetrahedra %%
from Delauney
tesselation. They then find the folding pathways that minimize the number
of times a ring forms before its thread is formed within a
single-sequence approximation: these indicate topologically-frustrated
pathways. 
As an interesting example in~\cite{NorcrossTS06}, strand IV of superoxide dismutase (SOD1) is highly
buried by parts of the Zn-binding loop, electrostatic loop, and
neighboring strands V and VII. In vitro folding
studies~\cite{KayatekinC08,NordlundA2009} show however that this problem is resolved
by Zn-binding after folding of the $\beta$-barrel, which is coupled
with structural formation of the Zn-binding and electrostatic loops
(loops IV and VII). The apo state is an energetically
stressed, metastable intermediate~\cite{DasA12pnas1}. 
In general, folding coupled to %%
ligand binding could remove topological frustration by inducing
unfrustrated pathways in the folding mechanism. 

Similar schematic ``average'' folding mechanisms 
as in Figures~\ref{fig:2abd_move}-\ref{fig:3mlg_move}, based on minimal
folding pathways, were proposed for the
complex Stevedore knotted protein $\alpha$-haloacid dehalogenase by
B\"{o}linger {\it et al}~\cite{BolingerD10}, based on folding simulation statistics of
G\={o} models.

Coarse-grained simulation studies of the reversible folder
YibK~\cite{MallamAL06} showed that non-native interactions between the
C-terminal end and residues towards the middle of the sequence were a
prerequisite for reliable folding to the trefoil knotted native
conformation~\cite{WallinS07}, the evolutionary origins of which were
supported by hydrophobicity and $\beta$-sheet propensity profiles of
the SpoU methyltransferase family. This suggests a new aspect of
evolutionary ``design'' involving selective non-native interactions, beyond the
generic role that non-native interactions may play in accelerating
folding rate~\cite{PlotkinSS01:prot,ClementiC04}. 
Low kinetic success rates $\sim 1-2\%$ in purely structure-based G\={o} simulations are
also seen in coarse-grained simulation studies of YibK~\cite{SulkowskaJI09} and
all-atom simulation studies of the small $\alpha/\beta$ knotted protein
MJ0366~\cite{NoelJK10}. In these studies by Onuchic and colleagues, a
``slip-knotting'' mechanism driven by native contacts is proposed, rather
than the ``plug'' mechanism in~\cite{WallinS07}, which is driven by
non-native contacts.  Both slip-knotting and plug mechanisms were
described by Mohazab and Plotkin as optimal un-crossing motions of
protein chains in~\cite{MohazabAR08:bj}. 
Such mechanisms may be facilitated by flexibility in the protein
backbone: highly conserved glycines in the hinge regions of both
knotted and slipknotted~\cite{KingNP07} proteins
modulate the knotted state of the corresponding subchain of
the protein~\cite{SulkowskaJI12}.
Further bioinformatic studies that investigate
evolutionary selection by strengthening critical native  or non-native
interactions in knotted proteins are an interesting topic of current
and future research. There 
is certainly a precedent of selection for native interactions that penalize
on-pathway intermediates in some proteins such as ribosomal
protein S6~\cite{PlotkinSS00:pnas,PlotkinSS02:Tjcp,LindbergM02}. 
Structural analysis of the deeply buried trefoil knot in acetohydroxy
acid isomeroreductase indicates 
swapping of secondary structural elements across replicated domains
likely arising from gene duplication~\cite{TaylorWR00}, which argues in
favor of knot formation driven by native interactions, through a 
mechanism apparently distinct from slipknotting. 

Lua and Grosberg have found that, due to enhanced
return probabilities originating from finite globule size along with
secondary structural preferences, protein chains have smaller degree
of interpenetration than collapsed random walks, and thus fewer knots
than would be expected for such collapsed random
walks~\cite{LuaRC06}, in spite of the fact that collapse dramatically
enhances the likelihood of knot formation~\cite{VirnauP05}, an effect
foreshadowed by the dramatic decrease in characteristic length for knot
formation %%
as solvent quality changes from
good to ideal (theta)~\cite{KoniarisK91,KoniarisK91-JCP}. 
It is still not definitively answered whether this statistical 
selection against knots in the protein universe is a cause or
consequence of the above size and structural preferences. 
Similarly, Mansfield~\cite{MansfieldML94,MansfieldML97} has suggested
that the polar nature of the N- and C- termini of the protein chain energetically penalize
processes that would result in the formation of knots. 

Conversely, some functional roles may benefit from the presence of
knotted topologies. Virnau and colleagues~\cite{VirnauP06} have suggested that the 
presence of complex knots in proteins involved in regulation of
ubiquitination and proteolysis serve a protective role against
incidental proteasome degradation, and as well, they observe evidence for
the modulation of function by alteration of an enzymatic binding site
through either the presence or absence of a knot in homologues of 
transcarbamylase. Phylogenetic analysis indicates that the presence of
a knot is most likely mediated by a 
single evolutionary event involving insertions of short segments in the
primary sequence~\cite{PotestioR10}. 

The interplay between sequence-determined energetics and chain
connectivity in the folding of proteins with complex or knotted
topologies is a topic of much current interest, despite the fact that
the number of proteins exhibiting knots or slipknots in their native
structures is relatively small. 
It will be interesting to see how evolution has optimized sequence or
facilitated protein-chaperone interactions to enable folding for these
``problem children'' of the proteome.

\section*{Acknowledgments}

We thank Atanu Das, Will Guest, and Stephen Toope for helpful and/or supportive
discussions. A.R.M. acknowledges Mohammad S.  Mashayekhi for
computational resource support.  We also acknowledge funding from the
Faculty of Graduate Studies 4YF Program at the University of
British Columbia at Vancouver, and the Natural Sciences and
Engineering Council for providing funding to defray publication page
fees.

\section*{Figure Legends}
\begin{figure}[!ht]
  \centering
  \caption[Transformation of a simple conformation with link size
  change shown]{{\bf Approximate minimal tranformation for a simple conformation
      pair, and the degree to which link length changes.} (a) Several intermediate conformations for a
    transformation (A--G proceeding along the color sequence red,
    green, yellow, cyan, magenta, gray, and blue) are shown. The
    step-size delta is shown. Note the step in which the first bead of
    the chain ($b_0$) is ``snapped'' into the final conformation
    because its distance to the destination is less than $\delta$
    (going from D to E). In the intermediate conformation F (Gray),
    beads 0 to 3 have reached their final locations and no longer
    move. Note also the link length violation of link 4 in
    conformation F, due to the approximation that ignores end point
    rotations, for this intermediate figure. A milder violation is
    observed when going from D (cyan) to E (magenta), since bead 1
    through $N$ all assume a step size of $\delta$ while bead 0 moved
    a step size $<\delta$. (b) Panel b shows a surface plot showing
    link length as a function of link number and step number during
    transformation. For the whole process, mean link length ${\bar
      \ell}$ is 0.98 units and standard deviation $\delta \ell$ is
    0.063.}
  \label{fig:method_delta}
\end{figure}
\begin{figure}[!ht]
  \centering
  \caption[Histogram for link length deviation]{{\bf Link length
      statistics for randomly generated transformation pairs.}
    Histogram of the average link length over the course of a
    transformation, for transformations
    between 200 randomly generated structures of 9 links and the
    (randomly generated) reference structure shown in the inset to the
    figure. %%
    The ``native'' or reference state is shown in
    the inset, along with several of the 200 initial states. %%
    For the ensemble of transformations shown, the ensemble
    average of the mean link length is 0.96. %%
  }
  \label{fig:link-deviation}
\end{figure}
\begin{figure}[!ht]
  \centering

  \caption[Crossing detection using projections]{{\bf Crossing
      detection using projections.} (a) A 3 link chain with its vertical projection. A crossing
    in the projection is shown with a green circle. The crossing in
    the projection occurs at points $s=0.29$ and $s=2.82$, where the
    chain is parametrized uniformly from 0 to 3. Since link 1 is under
    link 3 at the point of projection crossing, 0.29 will appear with
    a negative sign in the corresponding $\XX$ (eqn
    \ref{eq:eqcross1}). (b) The blue chain and the red chain have the
    exact same vertical projection, however their corresponding $\XX$
    matrices are different in sign, as given in
    Eq.~\ref{eq:eqcross2}. This indicates that the over-under sense
    has changed for the links whose projections are crossing. This in
    turn indicates that a true crossing has occurred when going from
    the red conformation to the blue conformation, as opposed to a
    series of conformations where the chain has navigated to
    conformations having the opposite crossing sense without passing
    through itself.}
  \label{fig:3linkchainX}

\end{figure}

\clearpage

\begin{figure}[!ht]
  \centering
  \caption[Two possible untangling transformations]{{\bf Two possible
      minimal uncrossing transformations.}Two possible untangling transformations. The top
    transformation involves twisting of the loop. The lower
    transformation involves a snake like movement of the vertical
    leg. A third one would involve moving the horizontal leg, in a similar
    snake-like fashion. Note that the moves represented here are not
    necessarily the most efficient ones in their topological class,
    but rather the most intuitive ones. There are transformations that
    are topologically equivalent but generally involve less total motion of
    the chain (see for example Figures~\ref{fig:loop_twist_all}(a),
    \ref{fig:loop_twist_all}(b)). 
  }
  \label{fig:simple_untangle}
\end{figure}
\begin{figure}[!ht]
  \centering
  \caption[Minimal untangling using knowledge of future
  crossings]{{\bf Accounting for history-dependence in minimal
      uncrossing transformations.} The minimal untangling movement in
    going from A to C  
    (through B') is less than the sum of the minimum untangling
    movements going from A to B and then from B to C }
  \label{fig:retrospect_matters}
\end{figure}
\begin{figure}[!ht]
  \centering
  \caption[Snapshots of a transformation with two crossings]{{\bf
      Snapshots of a transformation with two crossings.} A few snapshots during a transformation involving 2
    instances of chain crossing. The transformation occurs clockwise
    starting from initial configuration I and proceeding to final configuration F.}
  \label{fig:concrete_eg}
\end{figure}
\begin{figure}[!ht]
  \centering
  \caption[Leg substructure]{{\bf Identification of leg-uncrossing.} For the crossing points indicated by the green circles,
    two legs, colored blue and red, can be identified. Each leg starts at the
    crossing and terminates at an end.}
  \label{fig:leg_example}
\end{figure}

\begin{figure}[!ht]
  \centering
  \caption[Crossing substructures]{{\bf Crossing substructures.} (a) A
    single leg structure, (b) A loop structure, (c) An elbow
    structure.} 
  \label{fig:all_object}
\end{figure}

\begin{figure}[!ht]
  \centering
  \caption[Schematic illustration of the canonical leg movement]{{\bf
      Schematic illustration of the canonical leg movement.} Schematic
    illustration of the canonical leg movement, 
    either from left to right as in (a) or effectively its time
    reverse as in (b). Both transformations traverse the same
    distance. The transformation in (a) is equivalent to the ``plug''
    transformation analyzed in the context of folding simulations for
    trefoil knotted proteins~\cite{WallinS07}, while the transformation in (b) (see
    ref.~\cite{MohazabAR08:bj} for a detailed description of this
    transformation) is equivalent to the ``slipknotting''
    transformation more often observed in the folding of knotted
    proteins~\cite{NoelJK10}.}
  \label{fig:leg_movement}
\end{figure}
\begin{figure}[!ht]
  \centering
  \caption[A single leg movement to undo several crossings]{{\bf A
      single leg movement can undo several crossings.} One can reverse the over-under nature of all the crossings
    that have occurred on a leg, through a single leg movement.}
  \label{fig:leg_movement_extra_cross}
\end{figure}
\begin{figure}[!ht]
  \centering
  \label{fig:loop_twist_pinch}
  \caption[Topological loop twist]{{\bf Relation of minimal loop
      uncrossing to Reideneister type I moves.} (a)Reversing the over-under nature of a crossing through a
    topological loop twist: Reidemeister move type I. (b) By
    ``pinching'' the loop before the twist, the cost in distance for
    changing the crossing nature is reduced.}
  \label{fig:loop_twist_all}
\end{figure}
\begin{figure}[!ht]
  \centering
  \caption[Schematic of the canonical elbow move]{{\bf Schematic of
      the canonical elbow move.} Schematic of the canonical elbow move. From left to right}
  \label{fig:elbow_move}
\end{figure}
\begin{figure}[!ht]
  \centering
  \caption[Various crossing substructures in a simple example]{{\bf A
      simple example depicting various crossing substructures.} A chain with several self-crossing points before and after
    untangling. Various topological substructures that are discussed
    in the text are color coded. For the case of the legs (red and
    cyan) note that various other legs can be identified, for example
    a leg that starts at crossing 2 and ends at the red terminus. Here we color only the shortest legs from crossing 1 to the terminus as red, and crossing 2 to the opposite terminus as cyan.}
  \label{fig:operator_example}
\end{figure}
\begin{figure}[!ht]
  \centering
  \caption[An example
  (subset) tree of possible transformations for 
  a given crossing structure]{{\bf Illustration of the depth-first tree search algorithm for 
      the given crossing structure shown.} An example
    (subset) tree of possible transformations for 
    a given crossing structure. Accumulated distances are given inside
    the circles representing nodes of the tree; the non-crossing
    transformations and their corresponding distances are shown next
    to the branches of the tree.  The algorithm starts from the bottom
    node and proceeds to the top nodes, starting in this case along
    the right-most branch. The possible transformations to be
    considered as candidate minimal transformations are :
    $[L_C(3)R(2)R(1)]$, $[E(3,4)R(2)R(1)]$, $[R(2)L_N(1)]$ which then
    terminates because the accumulated distance exceeds the minimum so
    far of 25, and $[L_C(2)L_N(1)]$.}
  \label{fig:tree_of_poss}
\end{figure}
\begin{figure}[!ht]
  \centering
  \caption[Clustering of proteins depending on order parameter]{{\bf
      Clustering of protein classes depending on order parameter.} (A) Scatter plot of all proteins as a function of $\Dnx/N$
    and $LRO$. Knotted proteins are indicated as green circles and are
    clustered; unknotted proteins are clustered using with the black
    closed curve, and contain $\alpha$-helical proteins clustered in
    red, and mixed $\alpha$-$\beta$ proteins clustered in
    magenta. Beta proteins are indicated in blue.  Two and three state
    proteins are indicated as triangles and squares respectively.  LRO
    provides a strong discriminant agains $\alpha$ and mixed proteins,
    but not knotted and unknotted proteins, while $\Dnx/N$
    discriminates knotted from unknotted proteins, and moderately
    discriminates $\alpha$ proteins from mixed proteins.  (B) Scatter
    plot of all proteins as a function of $\Dnx$ and $N$. The
    rendering scheme for protein classes is the same as in panel
    (A). Kinetic 2-state folders are indicated by the black
    dashed curve.  Both $\Dnx$ and
    $N$ distinguish knotted from unknotted proteins, and 2-state from
    3-state proteins. By projecting $\alpha$ proteins and either mixed
    $\alpha$/$\beta$ or all-$\beta$ proteins onto each order
    parameter, one can see how $\Dnx$ can discriminate $\alpha$
    proteins from both mixed or $\beta$ proteins, while $N$
    cannot. This is despite the significant correlation between $\Dnx$
    and $N$. 
  }
  \label{fig:cluster}
\end{figure}
\begin{figure}[!ht]
  \centering
  \caption[Statistical significance for all order parameters in
  distinguishing between different classes of proteins]{{\bf Statistical significance for all order parameters in
      distinguishing between different classes of proteins.} The -log of
    the statistical significance is plotted as a function of pairs of
    protein classes, so that a higher number indicates better ability
    to distinguish between different classes. The blue horizontal line
    indicates a threshold of $5\%$ for statistical
    significance. 
  }
  \label{fig:class_distinguish}
\end{figure}
\begin{figure}[!ht]
  \centering
  \caption[Approximate scaling laws for MRSD and non-crossing distance
  per residue $\Dnx/N$, across proteins and for domains within a
  single protein]{{\bf Approximate scaling laws for MRSD and non-crossing distance
      per residue $\Dnx/N$, across proteins and for domains within a
      single protein.} 
    (A) MRSD (blue circles) as a function of chain
    length $N$ for our protein dataset. The slope of the best fit line on the
    log-log plot gives the power law scaling: $MRSD \sim
    N^{0.65}$. Non-crossing distance per residue $\Dnx/N$ (red circles) {\it vs.} $N$
    shows much larger scatter across native topologies, but follows an
    approximate scaling law $\Dnx/N \sim N^{1.33}$ which is
    superextensive, indicating an increasing importance of chain
    non-crossing per residue as system size is increased. At system sizes
    larger than $N \approx 3600$, even minimal motion is dominated by entanglement. 
    (B) Same quantities as in panel (A), but for the domains in proteins
    2A5E and 2HA8. The scaling laws are different than in panel (A), and
    show stronger chain-length dependence. For 2HA8, domains 1, 2 and 1-2
    together (the full protein) are considered; for 2A5E domains 1,2,3,4,
    1-2, 2-3, 3-4, 1-2-3, 2-3-4, and 1-2-3-4 (the full protein) are
    considered. Based on these scaling laws found by building up proteins
    from subdomains, at system sizes
    larger than $N \approx 400$, minimal pathways become
    entanglement-dominated.   
    (C) Schematic renderings of the domains, color-coded in 2HA8 (left)
    and 2A5E (right). 
  }
  \label{figscaling}
\end{figure}
\begin{figure}[!ht]
  \centering
  \label{fig:3mlg_render}
  \caption[Renderings of the three proteins whose minimal
  transformations we investigate in detail]{{\bf Schematic renderings of the three proteins whose minimal
      transformations we investigate in detail.} (A)  acyl-coenzyme A binding
    protein, PDB id 2ABD~\cite{AndersenKV93}, an all-$\alpha$ protein; (B) Src homology 3 (SH3) domain
    of phosphatidylinositol 3-kinase, PDB id 1PKS~\cite{KoyamaS93}, a largely
    $\beta$ protein; (C) The designed knotted protein 2ouf-knot, PDB id 3MLG~\cite{KingNP10}. 
  }
  \label{fig:3protsrender}
\end{figure}
\pagebreak
\begin{figure}[!ht]
  \centering
  \label{fig:2abd_LR}
  \caption[Bar plots for the noncrossing operations involved in
  minimal transformations, for the $\alpha$ protein 2ABD]{{\bf Bar
      plots for the uncrossing operations involved in minimal 
      transformations from an unfolded ensemble, for the $\alpha$ protein 2ABD.} 
    The sequence of noncrossing operations the transformation
    corresponding to a given pair of conformations is represented as a color-coded
    series of bars, with the sequence of moves going
    from right to left, and the length of the bar indicating the
    non-crossing distance undertaken by a particular move. 
    Red bars indicate N-terminal leg ($\LN$) uncrossing, green bars 
    indicate C-terminal leg ($\LC$) uncrossing, blue bars indicate
    Reidemeister ``pinch and twist'' loop uncrossing
    moves, and cyan bars indicate elbow uncrossing moves.
    The same set of 172 transformations is shown in panels A and B. Panel
    A sorts uncrossing transformations by rank ordering the following move
    types, largest to
    smallest: $\LN$, $\LC$, loop uncrossing, elbow move. Panel B 
    sorts moves by $\LC$, $\LN$, loop uncrossing, elbow move. 
    The scale bar underneath each panel indicates a distance of 100 in
    units of the link length. 
    The arrow in each panel denotes the ``most representative'' transformation, as
    defined in the text. 
  }
  \label{figtransformsalpha}
\end{figure}
\pagebreak
\begin{figure}[!ht]
  \centering
  \label{fig:1pks_LR}
  \caption[Bar plots of the noncrossing operations for the $\beta$-sheet
  protein 1PKS]{
    {\bf Bar plots of the uncrossing operations involved in minimal transformations
      for the $\beta$-sheet protein 1PKS.} See Figure~\ref{figtransformsalpha} and the text for
    more details.
    Red bars: $\LN$ uncrossing moves; green bars: $\LC$ uncrossing moves; Blue bars:
    loop uncrossing moves; Cyan bars: elbow uncrossing moves.
    The same set of 195 transformations is shown in panels A and B, sorted
    as in Figure~\ref{figtransformsalpha}. 
    The scale bar underneath each panel indicates a distance of 100 in
    units of the link length. 
  }
  \label{figtransformsbeta}
\end{figure}
\clearpage
\begin{figure}[!ht]
  \centering
  \label{fig:3mlg_LR}
  \caption[Bar plots of the noncrossing operations for the knotted
  protein 3MLG]{{\bf Bar plots of the uncrossing operations involved in the
      minimal transformations for the knotted
      protein 3MLG.} See Figure~\ref{figtransformsalpha} and the text for
    more details.
    Red bars: $\LN$ uncrossing moves; green bars: $\LC$ uncrossing moves; Blue bars:
    loop uncrossing moves; Cyan bars: elbow uncrossing moves.
    The same set of 90 transformations is shown in panels A and B, sorted
    as in Figure~\ref{figtransformsalpha}. 
    The scale bar underneath each panel indicates a distance of 100 in
    units of the link length.  
    The arrow in each panel denotes the ``most representative'' transformation, as
    defined in the text. 
    The transformation located 8 bars up from
    the bottom of Panel A requires both $\LN$ and $\LC$ moves, however
    both leg motions are very small. 
  } 
  \label{figtransformsknot}
\end{figure}
\clearpage
\begin{figure}[!ht]
  \centering
  \label{fig:cons_Other}
  \caption[Consensus histograms of the transformations described in
  Figures~\ref{figtransformsalpha}-\ref{figtransformsknot}]{
    {\bf Consensus histograms of the transformations described in
      Figures~\ref{figtransformsalpha}-\ref{figtransformsknot}.} See text for a description
    of the construction. Each bar represents the distance of a
    corresponding move type, N or C leg ($L_N$ or $L_C$), elbow $E$,
    or loop $R$. The order of the sequence of moves is taken from
    right to left along the x-axis.  An all-$\alpha$ protein (2ABD),
    an all-$\beta$ protein (1PKS), and a knotted protein (3MLG) are
    considered. 
    (A) Transformations with leg $L_N$ as the largest move. These encompass 
    15\% of the transformations those in the $\alpha$ protein, 16\% 
    of the transformations in the $\beta$ protein, and 73\% of the
    transformations for the knotted protein. 
    (B) Transformations with leg $L_C$ as the largest move, which encompass
    13\% of the $\alpha$ protein
    transformations, 54\% of $\beta$ protein transformations, and 
    18\% of knotted protein transformations. 
    (C) Transformations with either an elbow E or loop R as the largest
    move, which encompass
    71\% of the $\alpha$ protein
    transformations, 29\% of $\beta$ protein transformations, and 9\%
    of knotted protein transformations. 
  }
  \label{fig:consensus_moves}
\end{figure}

\begin{figure}[!ht]
  \centering
  \caption[Schematic of the most representative transformation for the
  $\alpha$ protein 2ABD]{{\bf Schematic of the most representative transformation for the
      $\alpha$ protein 2ABD.}}
  \label{fig:2abd_move}
\end{figure}

\begin{figure}
  \centering
  \caption{{\bf Schematic of the most representative transformation for the
      $\beta$ protein 1PKS.}}
  \label{fig:1pks_move}
\end{figure}

\begin{figure}[!ht]
  \centering
  \caption[Schematic of the most representative transformation for the
  knotted protein 3MLG]{{\bf Schematic of the most representative transformation for the
      knotted protein 3MLG.}}
  \label{fig:3mlg_move}
\end{figure}

\begin{figure}[!ht]
  \centering
  \caption[Noncrossing operations overlap between two
  transformations]{
    {\bf Overlap between minimal transformations.} Schematic diagram for the residues involved in uncrossing
    operations for two minimal transformations labelled by $\alpha$ and $\beta$,
    to illustrate the sequence overlap between transformations.}
  \label{fig:sequence_move_match}
\end{figure}

\begin{figure}[!ht]
  \centering
  \caption[Pathway overlap ($\Qab$) distributions for 3 proteins]{
    {\bf Distribution of pathway overlap between minimal transformations, for an $\alpha$, $\beta$, and knotted protein.}
    Pathway overlap ($\Qab$) distributions for the 3
    proteins in Figure~\ref{fig:3protsrender}, as defined by
    Equation~\eqref{eq:Qab}, operating on the transformations in
    Figure~\ref{figtransformsalpha}-\ref{figtransformsknot}.  (a) The pathway overlap distribution for
    the all-$\alpha$ protein 2ABD indicates a large contribution for
    $\Qab = 0$ (the peak height in the distribution is $\approx
    0.62$), indicating a diverse set of minimal transformations 
    fold the protein. The average $Q$ for these transformations is
    $0.18$.  (b) The pathway overlap distribution for the
    $\beta$-protein shows the emergence of a peak around $\Qab=1$,
    indicating partial restriction of folding pathways. The peak
    near $\Qab=0$ still carries more weight in the distribution. The
    average $Q = 0.45$. (c) The peak around $\Qab=1$ becomes
    dominant for the pathway overlap distribution of the 
    knotted protein, indicating the emergence of a dominant
    restricted minimal folding pathway. The average $Q = 0.62$.
  }
  \label{fig:Qdist}
\end{figure}

\clearpage
\section*{Tables}
\begin{table}[!ht]
  \centering
  \caption{{\bf Proteins studied in this paper.}}
  \begin{small}
    \begin{tabular}{|c|c|c|c|c|c|c|c|c|c|c|c|c|}
      \hline
      PDB    &x-State &2ndry str. &LRO	&RCO	&ACO	&MRSD	&RMSD	&$\langle\Dnx\rangle$	&$\langle\Dnx\rangle/N$ &$\langle\mathcal{D}\rangle(\times 10^3)$& $\langle\mathcal{D}\rangle/N$	&  $N$	\\[1mm]
      \hline
      1A6N	&3	&$\alpha$-helix	&1.4	&0.1	&14.0	&26.2	&29.2	&285	&1.9	&{4.24}	&28.1	&151	\\[1mm]
      1APS	&2	&Mixed	&4.2	&0.2	&21.8	&22.7	&25.4	&201	&2.1	&{2.43}	&24.8	&98	\\[1mm]
      1BDD	&2	&$\alpha$-helix	&0.9	&0.1	&5.2	&14.0	&14.9	&76.5	&1.3	&{0.91}	&15.2	&60	\\[1mm]
      1BNI	&3	&Mixed	&2.5	&0.1	&12.3	&20.8	&22.8	&209	&1.9	&{2.46}	&22.8	&108	\\[1mm]
      1CBI	&3	&$\beta$-sheet	&2.8	&0.1	&18.8	&25.1	&27.9	&286	&2.1	&{3.70}	&27.2	&136	\\[1mm]
      1CEI	&3	&$\alpha$-helix	&1.0	&0.1	&9.1	&16.7	&18.9	&71.4	&0.8	&{1.49}	&17.5	&85	\\[1mm]
      1CIS	&2	&Mixed	&3.3	&0.2	&10.8	&15.1	&16.8	&99.7	&1.5	&{1.10}	&16.6	&66	\\[1mm]
      1CSP	&2	&$\beta$-sheet	&3.0	&0.2	&11.0	&16.8	&18.4	&98.0	&1.5	&{1.23}	&18.3	&67	\\[1mm]
      1EAL	&3	&$\beta$-sheet	&2.5	&0.1	&15.7	&24.9	&27.9	&278	&2.2	&{3.44}	&27.1	&127	\\[1mm]
      1ENH	&2	&$\alpha$-helix	&0.4	&0.1	&7.4	&13.5	&14.9	&28.0	&0.5	&{0.76}	&14.1	&54	\\[1mm]
      1G6P	&2	&$\beta$-sheet	&3.8	&0.2	&11.7	&16.4	&18.0	&83.1	&1.3	&{1.17}	&17.7	&66	\\[1mm]
      1GXT	&3	&Mixed	&3.7	&0.2	&18.6	&21.1	&23.5	&148	&1.7	&{2.03}	&22.8	&89	\\[1mm]
      1HRC	&2	&$\alpha$-helix	&2.2	&0.1	&11.7	&19.6	&22.2	&126	&1.2	&{2.17}	&20.8	&104	\\[1mm]
      1IFC	&3	&$\beta$-sheet	&2.8	&0.1	&17.7	&25.1	&27.9	&284	&2.2	&{3.58}	&27.3	&131	\\[1mm]
      1IMQ	&2	&$\alpha$-helix	&1.7	&0.1	&10.4	&16.1	&17.9	&80.7	&0.9	&{1.46}	&17.0	&86	\\[1mm]
      1LMB	&2	&$\alpha$-helix	&1.1	&0.1	&7.1	&17.0	&18.6	&76.8	&0.9	&{1.55}	&17.9	&87	\\[1mm]
      1MJC	&2	&$\beta$-sheet	&3.0	&0.2	&11.0	&17.5	&19.2	&110	&1.6	&{1.32}	&19.1	&69	\\[1mm]
      1NYF	&2	&$\beta$-sheet	&2.8	&0.2	&10.6	&15.3	&17.0	&87.4	&1.5	&{0.97}	&16.8	&58	\\[1mm]
      1PBA	&2	&Mixed	&2.6	&0.1	&12.0	&18.9	&20.8	&156	&1.9	&{1.69}	&20.8	&81	\\[1mm]
      1PGB	&2	&Mixed	&2.1	&0.2	&9.7	&14.1	&15.7	&25.4	&0.5	&{0.81}	&14.5	&56	\\[1mm]
      1PKS	&2	&$\beta$-sheet	&3.8	&0.2	&15.2	&17.9	&20.2	&136	&1.8	&{1.50}	&19.7	&76	\\[1mm]
      1PSF	&3	&$\beta$-sheet	&2.8	&0.2	&11.7	&16.8	&19.4	&72.1	&1.0	&{1.23}	&17.8	&69	\\[1mm]
      1RA9	&3	&Mixed	&3.4	&0.1	&22.3	&25.5	&28.6	&402	&2.5	&{4.46}	&28.1	&159	\\[1mm]
      1RIS	&2	&Mixed	&3.0	&0.2	&18.4	&21.5	&23.9	&163	&1.7	&{2.25}	&23.2	&97	\\[1mm]
      1SHG	&2	&$\beta$-sheet	&3.0	&0.2	&10.9	&15.1	&16.7	&92.3	&1.6	&{0.95}	&16.7	&57	\\[1mm]
      1SRL	&2	&$\beta$-sheet	&3.1	&0.2	&11.0	&14.8	&16.3	&94.5	&1.7	&{0.92}	&16.5	&56	\\[1mm]
      1TIT	&3	&$\beta$-sheet	&4.1	&0.2	&15.8	&18.7	&20.8	&154	&1.7	&{1.82}	&20.4	&89	\\[1mm]
      1UBQ	&2	&Mixed	&2.4	&0.2	&11.5	&17.0	&18.9	&92.1	&1.2	&{1.39}	&18.2	&76	\\[1mm]
      1VII	&2	&$\alpha$-helix	&0.4	&0.1	&4.0	&8.1	&9.2	&4.1	&0.1	&{0.30}	&8.2	&36	\\[1mm]
      1WIT	&2	&$\beta$-sheet	&5.0	&0.2	&18.9	&20.4	&22.7	&168	&1.8	&{2.07}	&22.2	&93	\\[1mm]
      2A5E	&3	&Mixed	&2.6	&0.1	&8.3	&22.2	&23.9	&354	&2.3	&{3.82}	&24.5	&156	\\[1mm]
      2ABD	&2	&$\alpha$-helix	&2.3	&0.1	&12.0	&18.2	&20.0	&77.5	&0.9	&{1.65}	&19.1	&86	\\[1mm]
      2AIT	&2	&$\beta$-sheet	&4.1	&0.2	&14.4	&16.9	&18.7	&107	&1.5	&{1.36}	&18.3	&74	\\[1mm]
      2CI2	&2	&Mixed	&2.7	&0.2	&10.0	&15.1	&16.9	&78.3	&1.2	&{1.06}	&16.4	&65	\\[1mm]
      2CRO	&3	&$\alpha$-helix	&1.2	&0.1	&7.3	&14.0	&15.5	&37.3	&0.6	&{0.95}	&14.6	&65	\\[1mm]
      2HQI	&2	&Mixed	&4.3	&0.2	&13.6	&16.3	&18.4	&86.9	&1.2	&{1.26}	&17.5	&72	\\[1mm]
      2PDD	&2	&$\alpha$-helix	&1.0	&0.1	&4.8	&10.6	&11.5	&19.9	&0.5	&{0.48}	&11.0	&43	\\[1mm]
      2RN2	&3	&Mixed	&3.6	&0.1	&19.3	&27.7	&30.9	&521	&3.4	&{4.81}	&31.0	&155	\\[1mm]
      1O6D	&--${}^\dag$	&Knotted	&3.1	&0.1	&18.9	&26.2	&28.7	&515	&3.5	&{4.36}	&29.7	&147	\\[1mm]
      2HA8	&--${}^\dag$	&Knotted	&3.3	&0.1	&16.2	&25.7	&28.5	&671	&4.1	&{4.84}	&29.9	&162	\\[1mm]
      2K0A	&--${}^\dag$	&Knotted	&3.4	&0.1	&14.6	&22.4	&24.5	&369	&3.4	&{2.81}	&25.8	&109	\\[1mm]
      2EFV	&--${}^\dag$	&Knotted	&2.1	&0.2	&12.6	&20.0	&21.8	&147	&1.8	&{1.79}	&21.8	&82	\\[1mm]
      1NS5	&3	&Knotted	&2.9	&0.1	&18.2	&27.5	&30.4	&503	&3.3	&{4.71}	&30.8	&153	\\[1mm]
      1MXI	&3	&Knotted	&2.8	&0.1	&16.7	&26.1	&29.0	&643	&4.0	&{4.85}	&30.1	&161	\\[1mm]
      3MLG	&3	&Knotted	&1.2	&0.1	&21.4	&27.7	&30.8	&481	&2.8	&{5.16}	&30.5	&169	\\[1mm]
      \hline
    \end{tabular}
  \end{small}
  \caption*{${}^\dag$Data not available at present.}
  \label{tab:proteins_used}
\end{table}
\begin{landscape}
  \begin{table}[!ht]
    \centering
    \caption{{\bf Comparison of order parameters for various protein classes.}}
    \begin{tabular}[]{|c||c|c||c|c||c|c|}
      \hline \hline Class&{\bf INX}&${\bf P_{INX}}$&${\mbox{\bf LRO}}$&${\bf P_{LRO}}$&${\mbox{\bf RCO}}$&${\bf P_{RCO}}$\\
      \hline    \begin{tabular}[]{c}
        2-state folders\\   
        3-state folders\\
      \end{tabular}
      &    \begin{tabular}[]{c}
        7.55e-02\\
        8.25e-02 \\
      \end{tabular}&
      (3.93e-01)
      &    \begin{tabular}[]{c}
        2.7\\
        2.6 \\
      \end{tabular}&
      (9.46e-01)
      &    \begin{tabular}[]{c}
        1.58e-01\\
        1.31e-01 \\
      \end{tabular}&
      (5.07e-02)
      \\
      \hline    \begin{tabular}[]{c}
        $\alpha$-helix proteins   \\
        $\beta$-sheet proteins \\
        Mixed secondary structure\\
      \end{tabular}
      &    \begin{tabular}[]{c}
        5.21e-02\\
        9.04e-02 \\
        8.64e-02 \\
      \end{tabular}&
      \begin{tabular}[]{c}
        $\alpha\beta$:4.01e-05\\
        $\beta\mbox{\sc m}$:(5.71e-01) \\
        $\alpha\mbox{\sc m}$:5.44e-04\\
      \end{tabular}
      &    \begin{tabular}[]{c}
        1.2\\
        3.3 \\
        3.1 \\
      \end{tabular}&
      \begin{tabular}[]{c}
        $\alpha\beta$:7.40e-08\\
        $\beta\mbox{\sc m}$:(4.27e-01) \\
        $\alpha\mbox{\sc m}$:6.20e-07\\
      \end{tabular}
      &    \begin{tabular}[]{c}
        1.10e-01\\
        1.72e-01 \\
        1.56e-01 \\
      \end{tabular}&
      \begin{tabular}[]{c}
        $\alpha\beta$:3.34e-07\\
        $\beta\mbox{\sc m}$:(2.68e-01) \\
        $\alpha\mbox{\sc m}$:3.48e-03\\
      \end{tabular}
      \\
      \hline    \begin{tabular}[]{c}
        Unknotted  proteins \\ 
        knotted  proteins   \\
      \end{tabular}
      &    \begin{tabular}[]{c}
        7.79e-02\\
        1.30e-01\\
      \end{tabular}&
      1.48e-03
      &    \begin{tabular}[]{c}
        2.6\\
        2.7\\
      \end{tabular}&
      (9.20e-01)
      &    \begin{tabular}[]{c}
        1.49e-01\\
        1.24e-01\\
      \end{tabular}&
      1.49e-02
      \\ 
      \hline \hline Class&$\mbox{\bf ACO}$&${\bf P_{ACO}}$&$\mbox{\bf MRSD}$&${\bf P_{MRSD}}$&$\mbox{\bf RMSD}$&${\bf P_{RMSD}}$\\
      \hline    \begin{tabular}[]{c}
        2-state folders\\   
        3-state folders\\
      \end{tabular}
      &    \begin{tabular}[]{c}
        11.4\\
        14.7 \\
      \end{tabular}&
      4.50e-02
      &    \begin{tabular}[]{c}
        16.4\\
        21.9 \\
      \end{tabular}&
      5.89e-04
      &    \begin{tabular}[]{c}
        18.1\\
        24.4 \\
      \end{tabular}&
      4.88e-04
      \\
      \hline    \begin{tabular}[]{c}
        $\alpha$-helix proteins   \\
        $\beta$-sheet proteins \\
        Mixed secondary structure\\
      \end{tabular}
      &    \begin{tabular}[]{c}
        8.5\\
        13.9 \\
        14.5 \\
      \end{tabular}&
      \begin{tabular}[]{c}
        $\alpha\beta$:3.76e-04\\
        $\beta\mbox{\sc m}$:(7.08e-01) \\
        $\alpha\mbox{\sc m}$:1.62e-03\\
      \end{tabular}
      &    \begin{tabular}[]{c}
        15.8\\
        18.7 \\
        19.9 \\
      \end{tabular}&
      \begin{tabular}[]{c}
        $\alpha\beta$:(1.19e-01)\\
        $\beta\mbox{\sc m}$:(4.50e-01) \\
        $\alpha\mbox{\sc m}$:4.11e-02\\
      \end{tabular}
      &    \begin{tabular}[]{c}
        17.5\\
        20.8 \\
        22.1 \\
      \end{tabular}&
      \begin{tabular}[]{c}
        $\alpha\beta$:(1.14e-01)\\
        $\beta\mbox{\sc m}$:(4.73e-01) \\
        $\alpha\mbox{\sc m}$:4.16e-02\\
      \end{tabular}
      \\
      \hline    \begin{tabular}[]{c}
        Unknotted  proteins \\ 
        knotted  proteins   \\
      \end{tabular}
      &    \begin{tabular}[]{c}
        12.5\\
        16.9\\
      \end{tabular}&
      5.59e-03
      &    \begin{tabular}[]{c}
        18.3\\
        25.1\\
      \end{tabular}&
      1.79e-04
      &    \begin{tabular}[]{c}
        20.3\\
        27.7\\
      \end{tabular}&
      3.18e-04
      \\
      \hline \hline Class&$\boldsymbol{\Dnx}/{\bf N}$&${\bf P_{\boldsymbol{\Dnx}/N}}$&$\boldsymbol{\Dnx}$&${\bf P_{\boldsymbol{\Dnx}}}$&$\boldsymbol{\mathcal{D}}$&${\bf P_{\boldsymbol{D}}}$\\
      \hline    \begin{tabular}[]{c}
        2-state folders\\   
        3-state folders\\
      \end{tabular}
      &    \begin{tabular}[]{c}
        1.3\\
        1.9 \\
      \end{tabular}&
      1.71e-02
      &    \begin{tabular}[]{c}
        94.9\\
        238 \\
      \end{tabular}&
      3.30e-03
      &    \begin{tabular}[]{c}
        1309\\
        2924 \\
      \end{tabular}&
      8.06e-04
      \\
      \hline    \begin{tabular}[]{c}
        $\alpha$-helix proteins   \\
        $\beta$-sheet proteins \\
        Mixed secondary structure\\
      \end{tabular}
      &    \begin{tabular}[]{c}
        8.74e-01\\
        1.7 \\
        1.8 \\
      \end{tabular}&
      \begin{tabular}[]{c}
        $\alpha\beta$:1.88e-04\\
        $\beta\mbox{\sc m}$:(6.65e-01) \\
        $\alpha\mbox{\sc m}$:1.56e-03\\
      \end{tabular}
      &    \begin{tabular}[]{c}
        80.4\\
        146 \\
        195 \\
      \end{tabular}&
      \begin{tabular}[]{c}
        $\alpha\beta$:4.50e-02\\
        $\beta\mbox{\sc m}$:(2.99e-01) \\
        $\alpha\mbox{\sc m}$:2.30e-02\\
      \end{tabular}
      &    \begin{tabular}[]{c}
        1450\\
        1802 \\
        2274 \\
      \end{tabular}&
      \begin{tabular}[]{c}
        $\alpha\beta$:(4.14e-01)\\
        $\beta\mbox{\sc m}$:(3.10e-01) \\
        $\alpha\mbox{\sc m}$:(1.06e-01)\\
      \end{tabular}
      \\
      \hline    \begin{tabular}[]{c}
        Unknotted  proteins \\ 
        knotted  proteins   \\
      \end{tabular}
      &    \begin{tabular}[]{c}
        1.5\\
        3.3\\
      \end{tabular}&
      5.33e-04
      &    \begin{tabular}[]{c}
        144\\
        476\\
      \end{tabular}&
      2.05e-03
      &    \begin{tabular}[]{c}
        1862\\
        4074\\
      \end{tabular}&
      2.67e-03
      \\
      \hline
    \end{tabular}
    \begin{tabular}[]{|c||c|c||c|c|}
      \hline Class&$\boldsymbol{\mathcal{D}/N}$&${\bf P_{\boldsymbol{D}/N}}$&${\bf N}$&${\bf P_{N}}$\\
      \hline    \begin{tabular}[]{c}
        2-state folders\\   
        3-state folders\\
      \end{tabular}
      &    \begin{tabular}[]{c}
        17.6\\
        23.8 \\
      \end{tabular}&
      8.56e-04
      &    \begin{tabular}[]{c}
        71.3\\
        116 \\
      \end{tabular}&
      4.17e-04
      \\
      \hline    \begin{tabular}[]{c}
        $\alpha$-helix proteins   \\
        $\beta$-sheet proteins \\
        Mixed secondary structure\\
      \end{tabular}
      &    \begin{tabular}[]{c}
        16.7\\
        20.4 \\
        21.6 \\
      \end{tabular}&
      \begin{tabular}[]{c}
        $\alpha\beta$:(6.95e-02)\\
        $\beta\mbox{\sc m}$:(4.67e-01) \\
        $\alpha\mbox{\sc m}$:2.68e-02\\
      \end{tabular}
      &    \begin{tabular}[]{c}
        77.9\\
        83.4 \\
        98.3 \\
      \end{tabular}&
      \begin{tabular}[]{c}
        $\alpha\beta$:(6.57e-01)\\
        $\beta\mbox{\sc m}$:(2.49e-01) \\
        $\alpha\mbox{\sc m}$:(1.59e-01)\\
      \end{tabular}
      \\
      \hline    \begin{tabular}[]{c}
        Unknotted  proteins \\ 
        knotted  proteins   \\
      \end{tabular}
      &    \begin{tabular}[]{c}
        19.7\\
        28.4\\
      \end{tabular}&
      1.04e-04
      &    \begin{tabular}[]{c}
        86.9\\
        140\\
      \end{tabular}&
      3.54e-03
      \\
      \hline
    \end{tabular}
    \caption* {Order parameters for various
      classifications of proteins. The data set of 2- and 3-state
      folders is the same as the data set for $\alpha$-helical
      $\beta$-sheet and mixed proteins, and is given in
      table~\ref{tab:proteins_used}. This is also the same data set
      as the unknotted proteins. 
      Knotted proteins are separately classified, and not included as either
      2-state or 3-state proteins. 
      A discrimination is deemed statistically significant if the probability of the null
      hypothesis is less than $5\%$.  
    }  
    \label{tab:order_parameters_for_various_classes}
  \end{table}
\end{landscape}

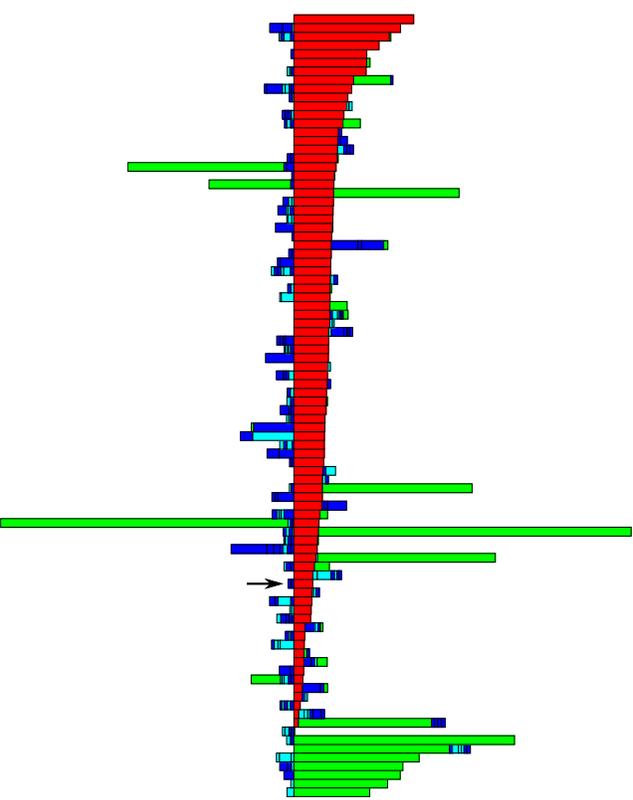
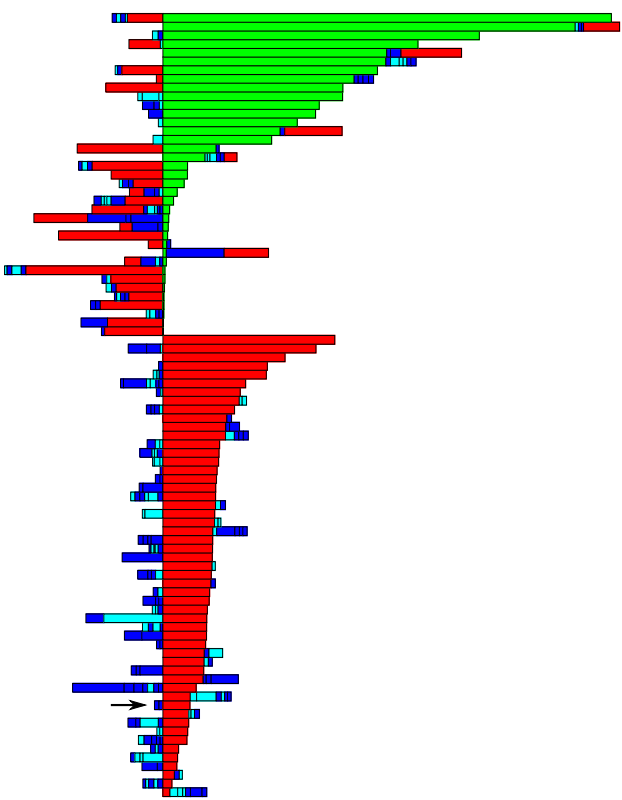
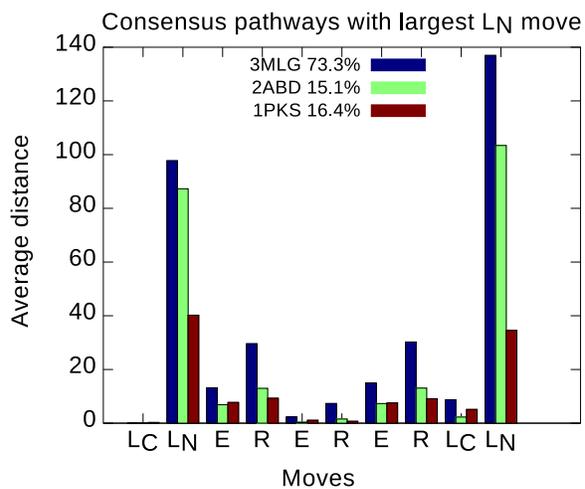
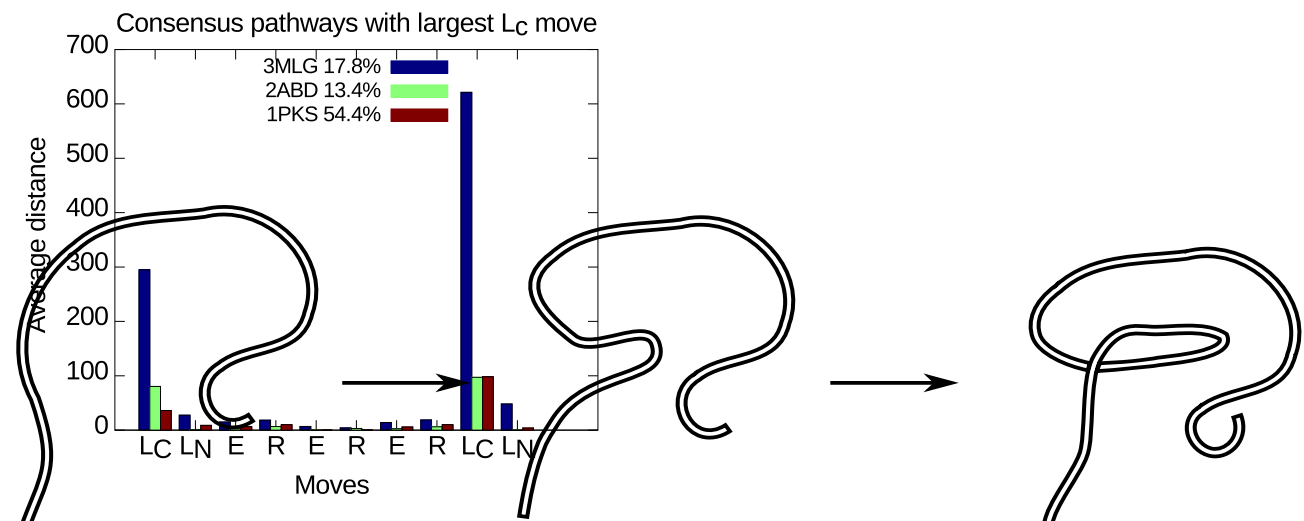
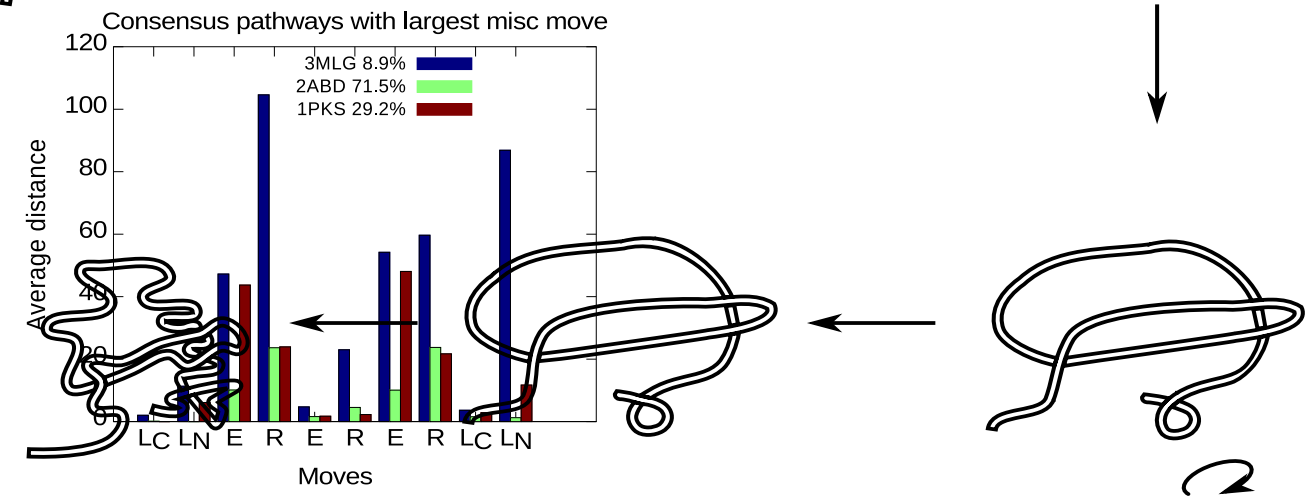
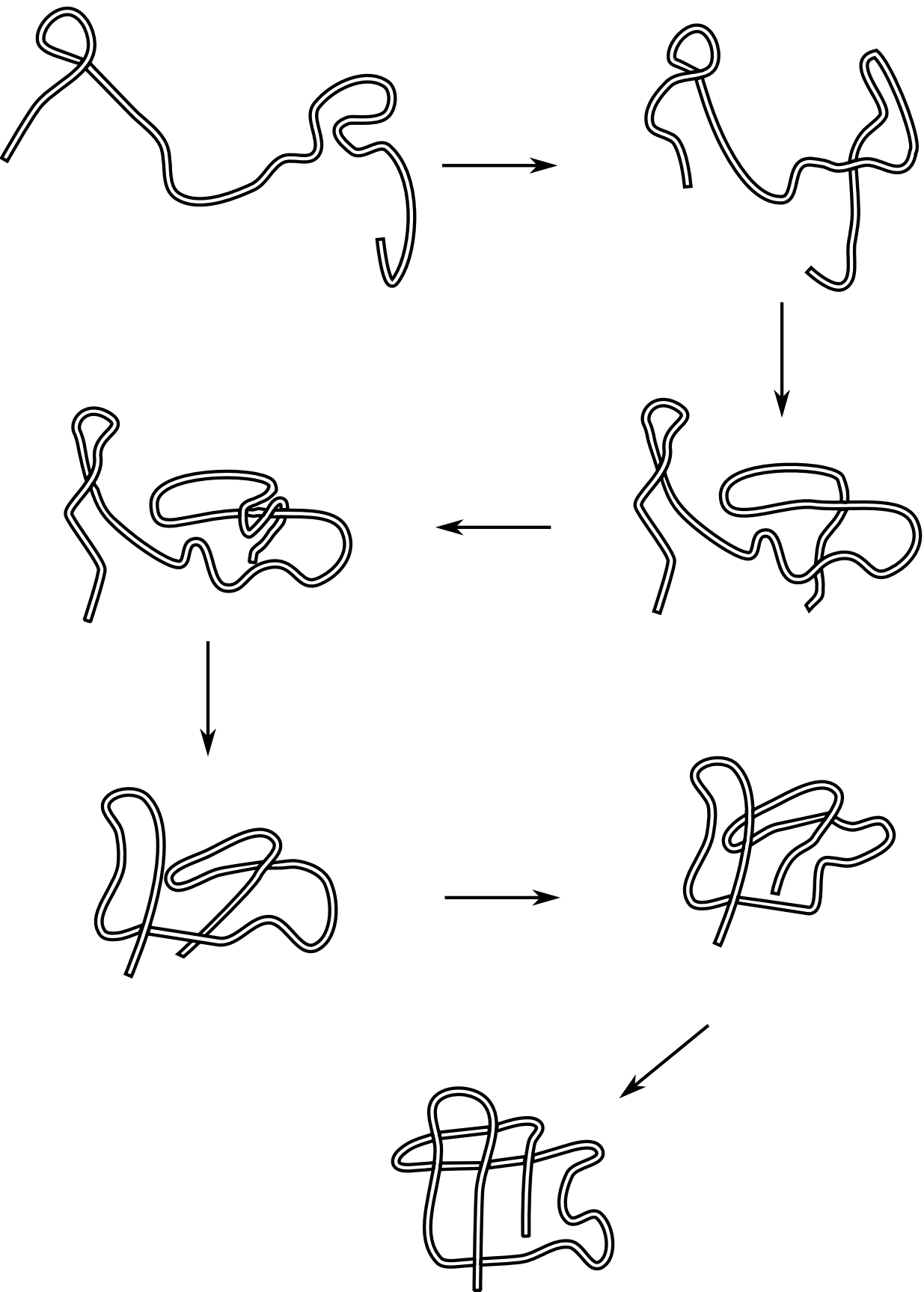
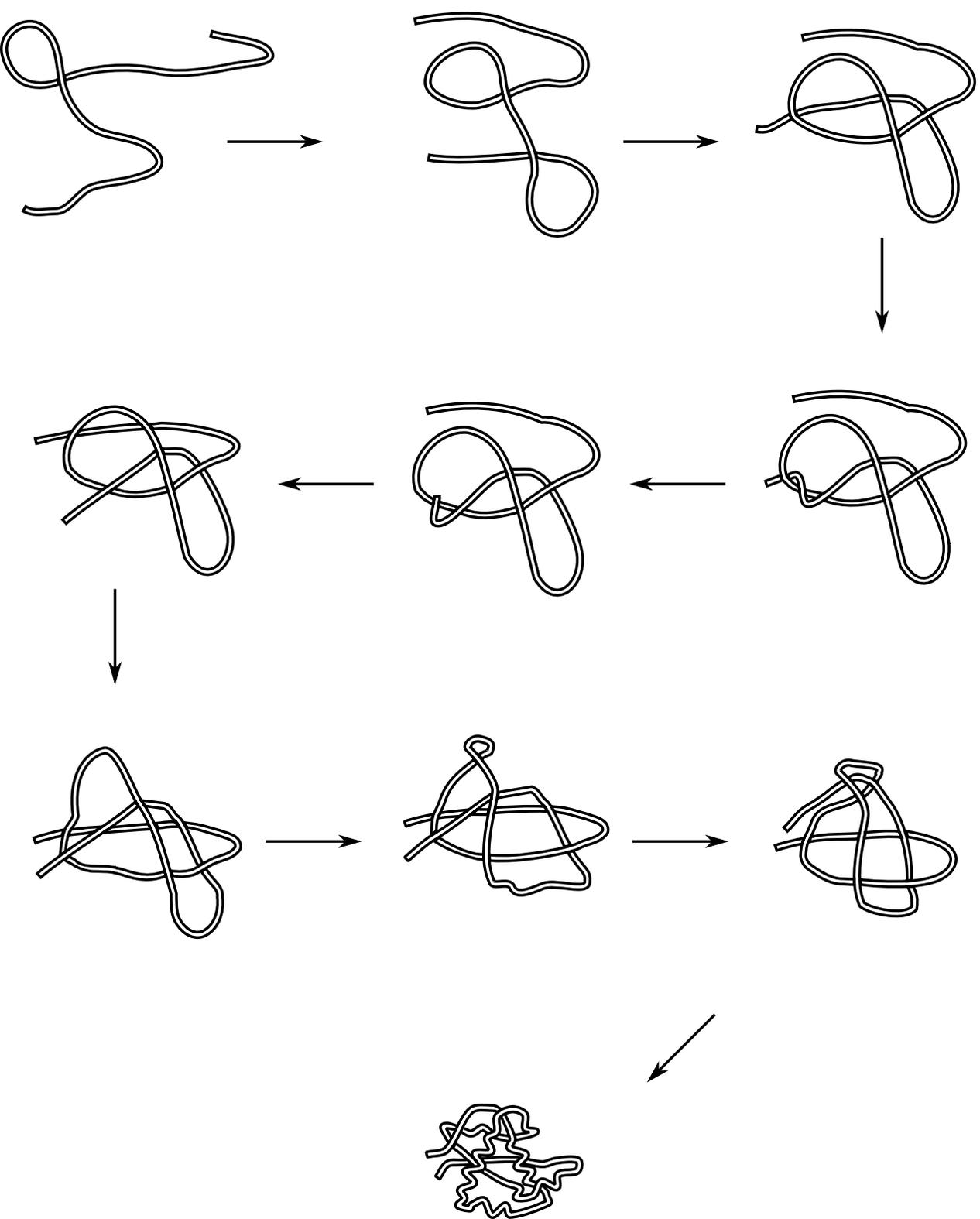
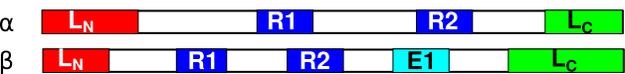
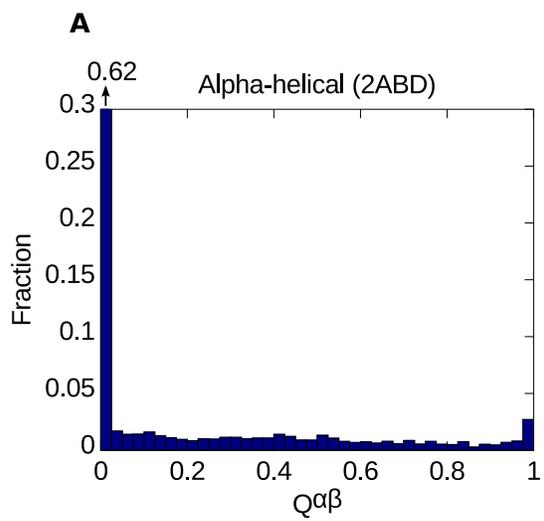
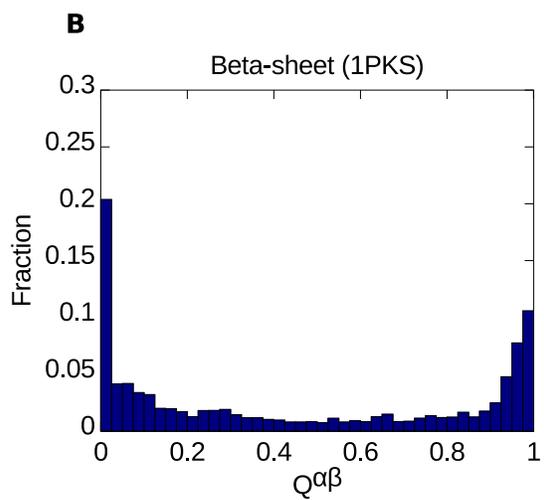
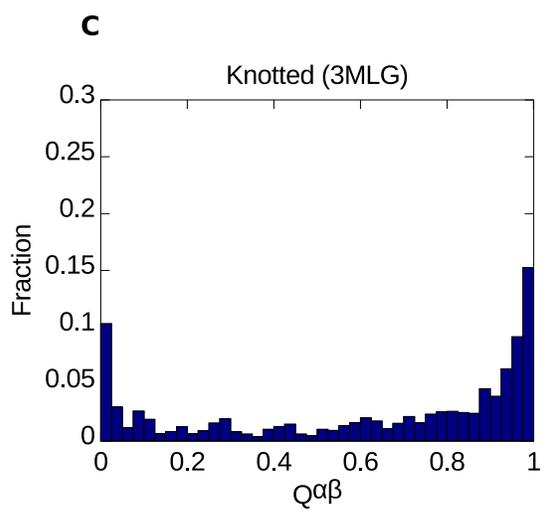
\end{document}